\documentclass[12pt,draftclsnofoot, onecolumn]{IEEEtran}

\usepackage{graphicx}
\usepackage{amsmath}
\usepackage{amssymb}
\usepackage[caption=false]{subfig}
\usepackage[noadjust]{cite}
\usepackage{float}
\usepackage{algorithm}
\usepackage{bibentry}
\usepackage{balance}
\usepackage{algorithm}
\usepackage{algorithmicx}
\usepackage{algpseudocode}
\usepackage{xcolor}
\usepackage{balance}
\usepackage{multirow}

\newtheorem{theorem}{Theorem}

\graphicspath{{img/}}
\newcommand*\dd{\mathop{}\!\mathrm{d}}

\begin{document}

\title{Low-Resolution Limited-Feedback NOMA for mmWave Communications}

\author{Yavuz Yap{\i}c{\i}, \.{I}smail G\"{u}ven\c{c}, and Huaiyu Dai 
\thanks{The authors are with the Department of Electrical and Computer Engineering, North Carolina State University, Raleigh, NC (e-mail:~\{yyapici,iguvenc,hdai\}@ncsu.edu).}
\thanks{This research is supported in part by NSF under the grant CNS-1618692.}}

\maketitle

\begin{abstract}
The spectrum-efficient millimeter-wave (mmWave) communications has recently attracted much attention as a viable solution to spectrum crunch problem. In this work, we propose a novel non-orthogonal multiple access (NOMA) framework, which makes use of the directional propagation characteristics of mmWave communications so as to improve the spectral efficiency through non-orthogonal signaling. In particular, we consider one-bit quantized angle information as a limited yet effective feedback scheme describing the channel quality of user equipment (UE) in mmWave bands. The UE pairs for NOMA transmission are then established using not only the one-bit distance feedback as a classical approach, but also the one-bit angle feedback. The proposed strategy is therefore referred to as two-bit NOMA. We also propose a novel hybrid strategy, called combined NOMA, for the circumstances with no UE pair through two-bit NOMA. Whenever no UE pair is available through any NOMA strategy, we resort to single user transmission (SUT) with proper UE selection schemes. The hybrid sum-rate performance is also analyzed thoroughly with the respective outage and rate expressions. The numerical results verify that the proposed strategy outperforms one-bit NOMA schemes with either angle- or distance-only feedback. 
\end{abstract}

\begin{IEEEkeywords}
Low-resolution limited-feedback, millimeter-wave (mmWave) communications, multiuser communications, non-orthogonal multiple access (NOMA). 
\end{IEEEkeywords}

\section{Introduction}\label{sec:intro}
The number mobile devices around the world has increased unprecedentedly over the last decade, which is reported to reach some 7.9 billion subscriptions in the first quarter of 2019 corresponding to an annual growth rate of $2\%$ \cite{ericsson2019mobrep}. To accommodate the wireless data demanded by this huge amount of mobile user equipment (UE), new communications bands and multiple access schemes have been considered for next-generation cellular networks in the scope of 5G and beyond. The use of millimeter-wave (mmWave) frequency bands has therefore become a promising solution to the spectrum crunch at the communications spectrum below $6\,\text{GHz}$, which occurs along with the new communications signals of much wider bandwidth \cite{Rappaport2016mmWave,Huang2018LowLat,Molisch2018MicMil}. In an effort to use the spectrum even more efficiently, non-orthogonal multiple access (NOMA) schemes are envisioned as the key technology for densely packed multiuser environments \cite{Dobre2017PowDom, Shao2018ComStu,Hanzo2018SurNom}.

The NOMA strategy has attracted much attention recently both for sub-6GHz and mmWave spectrum involving terrestrial and air-to-ground (A2G) communications \cite{Zhang2019UplCoo,Chen2019MulAnt,Alouini2019JoiTra,Assi2018NovCoo,Yu2019PlaPow}. As opposed to conventional orthogonal multiple access (OMA), this new strategy offers promising spectral efficiency since available resources are allocated to multiple UEs simultaneously~\cite{Dai2015NOMA,Ding2017AppNOMA,Dulaimi20185G}. This appealing performance, however, comes at the expense of additional link overhead. The NOMA transmitter allocates adequate power level to each UE based on their channel qualities, which requires feedback from each UE representing its channel, and in turn increases link overhead. It is therefore of significant importance to investigate the effective feedback strategies for NOMA transmission which aim at causing less overhead on the feedback link, and/or necessitating as low computational burden as possible at UEs (while computing requested feedback).

There are various works in the literature considering limited-feedback strategies for NOMA \cite{Ding2016DesMas,Schober2016OnThe,Jafarkhani2017DowNOMA,Lee2017NOMA,Ding2017RanBea,Yapici2019NOMAmmWDro_J,Yapici2019AngFee}. In particular, \cite{Ding2016DesMas} proposes a decomposition scheme for a massive multiple-input multiple-output (MIMO) NOMA transmission while representing the channel by a one-bit feedback. Similarly, \cite{Schober2016OnThe} analyzes the outage performance of NOMA under short- and long-term power constraints with the channel state information (CSI) represented by one-bit feedback. In \cite{Jafarkhani2017DowNOMA}, a special quantizer with variable-length encoding approach is proposed for NOMA transmission, which computes feedback bits directly from CSI. Performance of limited-feedback NOMA is studied in \cite{Lee2017NOMA} along with user scheduling and grouping strategies. \cite{Ding2017RanBea} considers a one-bit limited-feedback scheme for NOMA transmission with random beamforming, where the quantized bits are produced using the \textit{distance} of UEs to the transmitter. A distance-based one-bit scheme is considered in \cite{Yapici2019NOMAmmWDro_J} with the ultimate goal of improving the performance of unmanned aerial vehicle (UAV) communications employing NOMA transmission. In a follow-up work of \cite{Yapici2019AngFee}, angle-based one-bit feedback information is proposed for NOMA transmission, which leverages the directional propagation characteristic of mmWave spectrum.

In this study, we propose a novel low-resolution limited-feedback NOMA transmission strategy for mmWave communications. In particular, we exploit the angular sparsity of the mmWave channels, where many of the multipaths show up in a narrow angular window, referred to as angular spread, around the beamforming direction. The angular position of any UE, hence, provides some degree of information on the associated channel quality. Along with the distance between two ends of the transceiver, which describes the channel in a classical way through path loss \cite{Yapici2019NOMAmmWDro_J}, angle is therefore yet another limited-feedback information for mmWave communications representing the channel quality \cite{Yapici2019AngFee}. Based on both the distance and angle information, we design a low-resolution limited-feedback NOMA strategy with the contributions listed below:
\begin{itemize}
    \item[---] We consider \textit{both} the distance and angle information while forming limited feedback on the UE channel quality (required for NOMA transmission) to better represent UE channels in mmWave spectrum. This approach is therefore different from \cite{Yapici2019NOMAmmWDro_J} and \cite{Yapici2019AngFee} which respectively use distance and angle information \textit{separately} while constructing UE feedback. 
    \item[---] Furthermore, we consider the quantized versions of distance and angle information to cut down the feedback overhead. In particular, we introduce angle-based one-bit NOMA where the single feedback bit represents the \textit{angle} information. In addition, we also propose a novel \textit{two-bit NOMA} strategy which makes use of two feedback bits, where one of the bits is for the distance, and the other is for the angle.   
    \item[---] We also consider particular deployments where two-bit NOMA is not possible, and propose a novel \textit{combined NOMA} strategy to offer better spectral efficiency under such circumstances. We furthermore show how to choose the optimal user to allocate all the spectral resources, referred to as single user transmission (SUT), whenever one-bit, two-bit, or combined NOMA strategies are not feasible (i.e., no UE pair is possible). 
    \item[---]  The overall hybrid sum-rates involving any NOMA and SUT strategy are also analyzed rigorously with dedicated outage and rate expressions for each strategy. The numerical results verify the performance analysis, which also show that the proposed NOMA strategies are superior to not only conventional OMA but also the classical one-bit NOMA with the distance-based feedback only \cite{Poor2017RanBea,Yapici2019NOMAmmWDro_J}.     
\end{itemize}

To the best of our knowledge, this is the first time in the literature to consider not only the angle-based one-bit NOMA but also two-bit NOMA involving both the one-bit distance and angle information while constructing channel feedback for NOMA in mmWave frequency bands. The rest of the paper is organized as follows. Section~\ref{sec:system} introduces the system model under consideration. The two-bit and angle-based one-bit NOMA strategies are presented in Section~\ref{sec:twobit_noma}. The combined NOMA and SUT schemes are considered in Section~\ref{sec:combined_noma_and_sut}, where the respective hybrid rate derivation is presented in Section~\ref{sec:hybrid_rates}. The numerical results verifying the performance superiority of the proposed NOMA schemes are provided in Section~\ref{sec:results}, and the paper concludes with some final remarks in Section~\ref{sec:conclusion}.


\section{System Model} \label{sec:system}

\begin{figure}[!t]
	\centering
	\hspace*{-0.0in}
	\includegraphics[width=0.6\textwidth]{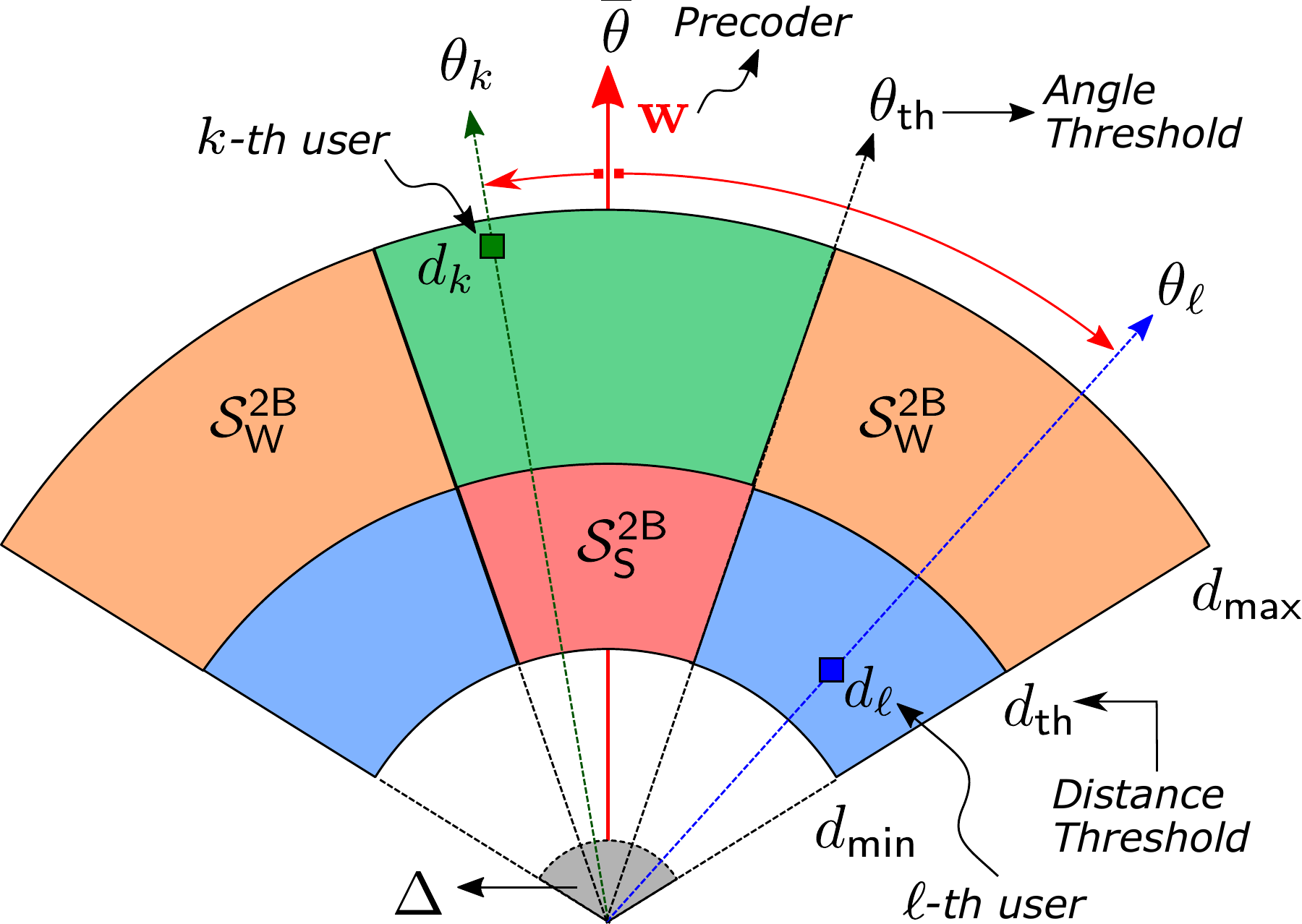}
	\caption{Polar representation of user region for two-bit NOMA in a multiuser mmWave communications scenario.}
	\label{fig:setting}
\end{figure}

We consider multiuser communications in mmWave frequency spectrum, where $K$ UEs with single antenna each are served simultaneously by a BS equipped with an $M$-element uniform linear array (ULA) antenna. The BS employs a stationary precoder $\textbf{w}$ to transmit UE messages after superposition coding (SC), and the UEs decode their messages following the successive interference cancellation (SIC) strategy. The overall transmission mechanism is therefore an example of NOMA, for which the UEs are distinguished in the power domain. We assume that all the UEs lie inside a user region which is represented in polar coordinates by the inner-radius $d_\mathsf{min}$, outer-radius $d_\mathsf{max}$, and center angle $\Delta$, as shown in Fig.~\ref{fig:setting}. Note that the user region can be adjusted through the parameters $\{d_\mathsf{min},d_\mathsf{max},\Delta\}$ to describe any communications environment of interest, which is exemplified in \cite{Yapici2019NOMAmmWDro_J, Yapici2019AngFee} considering a UAV-assisted wireless communications scenario intended for providing broadband coverage during temporary events (e.g., stadium concerts). 

We assume that the UEs are represented by the index set $\mathcal{N}_\mathsf{U} \,{=}\,  \{1,2,\ldots, K\}$, and are deployed randomly following a homogeneous Poisson point process (HPPP) with density $\lambda$. The number of UEs is therefore Poisson distributed with the probability mass function (PMF) $\mathsf{p}_K(k) \,{=}\, \frac{\mu^K e^{{-}\mu}}{K!}$ where $\mu \,{=}\, (d_\mathsf{max}^2{-}\,d_\mathsf{min}^2)\frac{\Delta}{2} \lambda$. The channel $\textbf{h}_k$ for the $k$-th UE is given as
\begin{align} \label{eq:channel_multipath}
\textbf{h}_k = \sqrt{M} \sum \limits_{p=1}^{N}  \frac{\alpha_{k,p}}{\sqrt{\mathsf{PL}\left(d_k\right)}}\,\textbf{a}(\theta_{k,p}),
\end{align} 
where $N$ is the number of multipath components, $\mathsf{PL}(d_k)$ is the path loss associated with the horizontal distance $d_k$ between the $k$-th UE and the BS, $\alpha_{k,p}$ is the complex gain of the $p$-th multipath which follows a zero-mean complex Gaussian distribution with variance $\sigma^2$, i.e., $\mathcal{CN}(0,\sigma^2)$, and $\theta_{k,p}$ is the angle-of-departure (AoD) of the $p$-th multipath. Furthermore, the $m$-th element of the steering vector $\textbf{a}(\theta_{k,p})$ is given as
\begin{align}
\Big[ \textbf{a}(\theta_{k,p}) \Big]_m =  \frac{1}{\sqrt{M}} {\rm exp} \left\lbrace {-}j2\pi \frac{d}{\lambda} \left( m{-}1\right) \cos\left( \theta_{k,p} \right) \right\rbrace,
\end{align}
for $m \,{=}\, 1,\dots,M$, where $d$ is the antenna element spacing of ULA, and $\lambda$ is the wavelength of the carrier frequency.

\section{Two-Bit NOMA} \label{sec:twobit_noma}

In this section, we describe the NOMA strategy with coding and decoding schemes under consideration, present the low-resolution limited-feedback schemes with distance and angle information, and introduce one-bit and two-bit NOMA strategies.  

\subsection{NOMA Transmission and Message Decoding} 
We assume that the UEs in $\mathcal{N}_\mathsf{U}$ are indexed from the best to the worst channel based on the feedback transmitted back to the BS on the UE channel qualities. We further assume that a subset of all the UEs, which is denoted by $\mathcal{N}_\mathsf{N}$ involving $K_\mathsf{N}$ elements, take place in NOMA transmission such that $\mathcal{N}_\mathsf{N} \,{\displaystyle\subset}\, \mathcal{N}_\mathsf{U}$. Following SC scheme, the desired UE messages was gathered to produce the transmit signal, which is given as
\begin{align}\label{eq:signal_transmitted}
\textbf{x} = \sqrt{\mathsf{P}_\mathsf{Tx}} \, \textbf{w}\sum\nolimits_{k\in \mathcal{N}_\mathsf{N}} \beta_k s_k, 
\end{align} 
where $\mathsf{P}_\mathsf{Tx}$ is the total downlink transmit power, $s_k$ is the unit-energy message symbol for the $k$-th UE, $\textbf{w}\,{\in}\,\mathbb{C}^{M{\times}1}$ is the precoder vector, and $\beta_k$ is the power allocation coefficient for the $k$-th UE such that $\sum_{k\in \mathcal{N}_\mathsf{N}} \beta_k^2\,{=}\,1$. Note that although optimal power allocation policy is not the main focus of this study, we assume that each UE is allocated a certain amount of power which is \textit{inversely} proportional to its channel quality to yield sufficient decoding performance. This power allocation policy yields $\beta_j\,{\leq}\,\beta_i$ for $\forall \, j\,{\leq}\,i$ with $i,j \,{\in}\, \mathcal{N}_\mathsf{N}$ since the UEs are ordered from the best channel quality to the worst.

The received signal at the $k$-th UE is given as
\begin{align} \label{eq:received_signal}
y_{k}= \textbf{h}_{k}^{\rm H} \textbf{x} +  v_k = \sqrt{\mathsf{P}_\mathsf{Tx}}\textbf{h}_{k}^{\rm H} \textbf{w}\sum \nolimits_{k\in \mathcal{N}_\mathsf{N}} \beta_k s_{k} + v_k,
\end{align} 
where $v_k$ is the observation noise being complex Gaussian with zero mean and variance $N_0$, i.e., $\mathcal{CN}(0,N_0)$, and the transmit signal-to-noise ratio (SNR) is therefore $\rho \,{=}\, \mathsf{P}_\mathsf{Tx} / N_0$. We further assume that the $k$-th UE has its own quality-of-service (QoS) based target rate $\overline{\mathsf{R}}_k$, and its message $s_k$ is accurately decoded whenever the respective instantaneous data rate exceeds $\overline{\mathsf{R}}_k$.

Exploiting the fact that the UEs with weak channel quality are allocated more power, each UE first decodes the relatively weaker UEs' messages in a sequential fashion starting from the weakest UE. While a message is being decoded, the interfering messages of all the other UEs (having stronger channel quality) is treated as noise. Adopting SIC strategy, each decoded message is then subtracted from the received signal (prior to the decoding of the next weaker UE's message), and the desired message of that UE is decoded once all the weaker UEs' messages are decoded and cancelled perfectly.

Assuming that all the interfering messages of the UEs weaker than $m$-th UE are decoded accurately at the $k$-th UE with $m\,{\geq}\,k$, the respective signal-to-interference-plus-noise ratio (SINR) while decoding the $m$-th UE message is given as 
\begin{align} \label{eq:sinr}
\mathsf{SINR}_{m{\rightarrow}k} = \frac{\mathsf{P}_\mathsf{Tx}|\textbf{h}_{k}^{\rm H}\textbf{w}|^2 \beta_{m}^2}{ \sum \limits_{l < m,\, l \in \mathcal{N}_\mathsf{N}} \mathsf{P}_\mathsf{Tx} |\textbf{h}_{k}^{\rm H}\textbf{w}|^2 \beta_{l}^2 + N_0},
\end{align} 
which also represents the SINR of $k$-th UE while decoding its own message for $m\,{=}\,k$. Note that \eqref{eq:sinr} yields $\frac{\mathsf{P}_\mathsf{Tx}}{N_0}|\textbf{h}_{k}^{\rm H}\textbf{w}|^2 \beta_{k}^2$ when the $k$-th UE being the strongest one is decoding its own message since no possible $l$ index of the summation in the denominator satisfies $l \,{<}\, k$. Defining the instantaneous rate associated with the $k$-th UE while decoding $m$-th UE message as $\mathsf{R}_{m{\rightarrow}k}\,{=}\,\log_2 \left( 1\,{+}\,\mathsf{SINR}_{m{\rightarrow}k}\right)$, the respective outage probability for a given $K \,{\geq}\, K_\mathsf{N}$ is given as
\begin{align} 
\mathsf{P}_{k|K}^\mathsf{o, NOMA} &= 1 - \mathsf{Pr} \left( \raisebox{0.05in}{$\displaystyle\bigcap\limits_{l \geq k,\, l \in \mathcal{N}_\mathsf{N}}$} \mathsf{R}_{l{\rightarrow}k} > \overline{\mathsf{R}}_{l} \right), \label{eq:outage_prob_noma_1} \\
&= 1 - \mathsf{Pr} \left( \raisebox{0.05in}{$\displaystyle\bigcap\limits_{l \geq k,\, l \in \mathcal{N}_\mathsf{N}}$} \mathsf{SINR}_{l{\rightarrow}k} > \epsilon_{l} \right), \label{eq:outage_prob_noma_2} 
\end{align} 
where $\epsilon_k\,{=}\,2^{\overline{R}_k}\,{-}\,1$. Note that \eqref{eq:outage_prob_noma_1} ensures that the $k$-th UE decodes its message successfully only if it can decode all the messages of the relatively weaker UEs. We therefore ignore \textit{error propagation} due to the unsuccessful decoding of interfering message of any weaker UE, which would be an interesting future topic to investigate. 

\subsection{Low-Resolution Limited Feedback} \label{sec:noma_feedback}
As discussed in the previous section, the power allocation policy dictates that each UE is allocated a certain amount of power which is inversely proportional to its channel quality. The NOMA transmitter therefore requires each UE to feedback an appropriate information describing its channel quality. Considering \eqref{eq:sinr}, the term $|\textbf{h}_{k}^{\rm H}\textbf{w}|^2$ completely describes the channel quality, and, hence, is referred to as \textit{effective channel gain}. Assuming that $\textbf{w} \,{=}\, \textbf{a}(\overline{\theta})$ with $\overline{\theta}$ being the AoD of the precoder vector, and incorporating the channel model in \eqref{eq:channel_multipath}, the effective channel gain is obtained as
\begin{align}\label{eq:effective_channel_gain}
|\textbf{h}_k^{\rm H}\textbf{w}|^2 &= \sum \limits _{p=1}^{N_{\rm P}} \frac{|\alpha_{k,p}|^2 }{\mathsf{PL}\left(d_k\right)} 
\underbrace{\left| \frac{ \sin \left( \frac{\pi M(\sin \overline{\theta}-\sin \theta_{k,p})}{2} \right)}{  M\sin \left( \frac{\pi (\sin \overline{\theta}-\sin \theta_{k,p})}{2} \right)}\right|^2}_{{\mathsf{F}_M(\overline{\theta},\theta_{k,p})}},
\end{align}
where $\mathsf{F}_M(\overline{\theta},\theta_{k,p})$ is the Fej\'er Kernel function \cite{Poor2017RanBea}. In the rest of the paper we consider only the impact of line-of-sight (LoS) path (i.e., $N_{\rm P} \,{=}\, 1$), since power of the LoS link is reported to be as large as $20\,\text{dB}$ above the other multipaths for a typical mmWave communications scenario \cite{Rappaport2012AngDep, Lee2016RanDir, Poor2017RanBea}. We therefore drop the subscripts representing individual multipaths from now on.  

Although the effective channel gain in \eqref{eq:effective_channel_gain} describes the channel quality completely, it becomes computationally expensive to estimate this term. Note that $|\textbf{h}_{k}^{\rm H}\textbf{w}|^2$ experiences rapid fluctuations over time due to the term $|\alpha_{k}|^2$ representing small-scale fading, and therefore requires sophisticated algorithms either to track the fading coefficient, or to frequently estimate it from the scratch \cite{Yapici2018LowComp}. In addition, large number of transmit antennas makes the candidate estimation and tracking algorithms computationally even more expensive. Note that the effective channel gain is also a function of the distance $d_k$ and the angle $\theta_{k}$, both of which vary slowly as compared to the small-scale fading coefficient $\alpha_{k}$, which is behind the rapid channel fluctuations. The angle and distance information are therefore considered \textit{separately} in \cite{Yapici2019AngFee} as \textit{limited feedback} for NOMA transmission, which turns out to be powerful alternatives of sending the complete effective channel gain back to the transmitter.  

In this work, we propose to use the \textit{low-resolution} (i.e., quantized) versions of the distance and angle information not only \textit{individually} (i.e., either the angle or distance) but also \textit{jointly} (i.e., both the angle and distance). To this end, we introduce one-bit quantized feedback information, which corresponds to thresholding the distance and angle using adequate threshold levels $d_\mathsf{th}$ and $\theta_\mathsf{th}$, respectively. Each UE therefore computes one-bit feedback for both the distance and angle information, and sends this low-resolution information back to the NOMA transmitter as representatives of its channel quality. 

\subsection{One-Bit and Two-Bit NOMA} \label{sec:one-bit-two-bit-noma}

The feedback information, which involves two bits for each UE, is then processed at the NOMA transmitter to split the UEs into two groups based on the channel qualities being strong or weak. When the NOMA transmitter employs both the feedback bits, which is referred to as \textit{two-bit NOMA}, the strong and weak UE groups are defined, respectively, as 
\begin{align}
    \mathcal{S}_\mathsf{S}^\mathsf{2B} &= \left\lbrace k \in \mathcal{N}_\mathsf{U} \,|\, | \overline{\theta} \,{-}\, \theta_k| \leq \theta_\mathsf{th} , d_k \leq d_\mathsf{th} \right\rbrace , \label{eq:set_strong_2bit}\\
    \mathcal{S}_\mathsf{W}^\mathsf{2B} &= \left\lbrace k  \in \mathcal{N}_\mathsf{U} \,|\, | \overline{\theta} \,{-}\, \theta_k| \geq \theta_\mathsf{th} , d_k \geq d_\mathsf{th} \right\rbrace ,\label{eq:set_weak_2bit}
\end{align}
which exploit the dependency of the effective channel gain on the distance and angle information, as provided by \eqref{eq:effective_channel_gain}. Similarly, the strong and weak UE groups based on angle information only are described, respectively, as
\begin{align}
    \mathcal{S}_\mathsf{S}^\mathsf{A} &= \left\lbrace k \in \mathcal{N}_\mathsf{U}  \,|\, | \overline{\theta} \,{-}\, \theta_k| \leq \theta_\mathsf{th} \right\rbrace, \label{eq:set_strong_angle}\\ 
    \mathcal{S}_\mathsf{W}^\mathsf{A}  &= \left\lbrace k \in \mathcal{N}_\mathsf{U}  \,|\, | \overline{\theta} \,{-}\, \theta_k| \geq \theta_\mathsf{th} \right\rbrace ,\label{eq:set_weak_angle}
\end{align}
and those for distance information only are
\begin{align}
    \!\!\!\mathcal{S}_\mathsf{S}^\mathsf{D} &\,{=} \left\lbrace k \in \mathcal{N}_\mathsf{U}  \,|\, d_k \leq d_\mathsf{th} \right\rbrace, \, 
    \mathcal{S}_\mathsf{W}^\mathsf{D} \,{=} \left\lbrace k \in \mathcal{N}_\mathsf{U}  \,|\, d_k \geq d_\mathsf{th} \right\rbrace , \label{eq:set_strong_weak_distance}
\end{align}
where we refer to these angle- or distance-only feedback schemes as \textit{one-bit NOMA}. In the rest of the paper, we use a similar convention where the subscripts/superscripts $\mathsf{2B}$, $\mathsf{A}$, and $\mathsf{D}$ stand for two-bit NOMA, one-bit NOMA with angle-only feedback, and one-bit NOMA with distance-only feedback, respectively, while $\mathsf{S}$ and $\mathsf{W}$ represent the channel quality being strong and weak, respectively.    

Assuming a practical transmission scheme of two UEs being served simultaneously, both the two-bit and one-bit NOMA transmitters pair UEs using these groups such that the strong (weak) UE with the index $k_\mathsf{S}$ ($k_\mathsf{W}$) is picked up from $\mathcal{S}_\mathsf{S}^\mathsf{t}$ ($\mathcal{S}_\mathsf{W}^\mathsf{t}$) arbitrarily, where $\mathsf{t} \,{\in}\, \{\mathsf{2B,A,D}\}$ and $\mathcal{N}_\mathsf{N} \,{=}\, \{k_\mathsf{S}, k_\mathsf{W}\}$ with $k_\mathsf{S} \,{<}\, k_\mathsf{W}$. By this way, each pair consists of UEs with sufficiently \textit{distinctive} channel qualities, which ensures the promised performance of NOMA. In this two-user NOMA scheme, outage probability of the strong UE is given using \eqref{eq:outage_prob_noma_1} as follows
\begin{align}
    \mathsf{P}_\mathsf{S,t}^\mathsf{o,NOMA} = 1 - \mathsf{Pr} \Big(  \mathsf{SINR}_{k_\mathsf{W}{\rightarrow}k_\mathsf{S}} > \epsilon_{\mathsf{W}}, \mathsf{SINR}_{k_\mathsf{S}{\rightarrow}k_\mathsf{S}} > \epsilon_{\mathsf{S}} \Big), \label{eq:outage_noma_strong} 
\end{align}
where $\epsilon_\mathsf{s}\,{=}\,2^{\overline{\mathsf{R}}_\mathsf{s}}\,{-}\,1$ with $\mathsf{s} \,{\in}\, \{\mathsf{S,W}\}$, $\overline{\mathsf{R}}_\mathsf{S}$ and $\overline{\mathsf{R}}_\mathsf{W}$ are the target rates for the strong and weak UEs, respectively, and $\mathsf{t} \,{\in}\, \{\mathsf{2B,A,D}\}$. Similarly, the outage probability for the weak NOMA UE is 
\begin{align}
    \mathsf{P}_\mathsf{W,t}^\mathsf{o,NOMA} = 1 - \mathsf{Pr} \Big(  \mathsf{SINR}_{k_\mathsf{W}{\rightarrow}k_\mathsf{W}} > \epsilon_{\mathsf{W}} \Big). \label{eq:outage_noma_weak} 
\end{align}

We would like to note that although distance-based one-bit NOMA is considered in \cite{Poor2017RanBea}, there is neither angle-based one-bit NOMA nor two-bit NOMA strategies available in the literature for conventional radio-frequency (RF) communications. Considering the directional transmission feature of the mmWave communications, the angle information, however, plays a crucial role in describing the channel quality of mmWave links, and is definitely helpful for NOMA transmission. The numerical results of Section~\ref{sec:results} verify the superiority of the use of angle information in both one-bit and two-bit NOMA for mmWave communications. Note also that more than two UEs can also be scheduled simultaneously within this framework, but at the expense of losing practicality and degraded performance of SIC decoding due to the worse UE separation in terms of channel qualities. 

\section{SUT and Combined NOMA} 
\label{sec:combined_noma_and_sut}

In this section, we describe the SUT and combined NOMA strategies aiming at further improving the spectral efficiency based on the form of the available limited feedback. 

\subsection{Single User Transmission} \label{sec:sut_scheme}

Depending on the NOMA user pairing strategy and the feedback mechanism (on the UE channel qualities), it is not always possible to form a NOMA user group involving $K_\mathsf{N}$ UEs, where $K_\mathsf{N} \,{=}\, 2$ for one-bit and two-bit NOMA. We therefore assume a hybrid transmission strategy such that the time-frequency resources and transmit power are fully allocated to a single user whenever NOMA transmission is not possible, and referred to this particular scheme as \textit{single user transmission (SUT)}. 

When it is not possible to set up a UE pair, and, hence, NOMA is not feasible, the SUT scheme should choose the \textit{best} UE to schedule. Indeed, under any circumstances, the best UE to schedule during SUT is the one with the best channel quality. However, since the exact channel quality of the UEs is not available completely at the transmitter, we need to determine the best UE based on the available low-resolution limited feedback.

Assuming that the two-bit feedback information is available at the transmitter, the best UE to schedule can be found by searching the UE groups using the order $\mathcal{S}_\mathsf{S}^\mathsf{2B} \rightarrow \mathcal{S}_\mathsf{W}^\mathsf{2B}$.
In this mechanism, the transmitter randomly picks up a UE from $\mathcal{S}_\mathsf{S}^\mathsf{2B}$ provided that this group is not empty, and otherwise repeat the same procedure for $\mathcal{S}_\mathsf{W}^\mathsf{2B}$. Similarly, when the angle-only (distance-only) information is available at the transmitter, the best UE for SUT can be found by searching the groups following the order $\mathcal{S}_\mathsf{S}^\mathsf{A} \rightarrow \mathcal{S}_\mathsf{W}^\mathsf{A}$ ($\mathcal{S}_\mathsf{S}^\mathsf{D} \rightarrow \mathcal{S}_\mathsf{W}^\mathsf{D}$). 

Note that the use of one-bit feedback ends up with either NOMA or SUT for $K \,{\geq}\, 2$. On the other hand, two-bit feedback might conclude with a no-transmission situation (other than NOMA and SUT) since $\mathcal{S}_\mathsf{S}^\mathsf{2B}$ and $\mathcal{S}_\mathsf{W}^\mathsf{2B}$ do not span the whole UE region, as sketched in Fig.~\ref{fig:setting}. Assuming that the SUT mechanism schedules a UE with the index $k_\mathsf{SUT} \,{\in}\, \mathcal{S}_\mathsf{s}^\mathsf{t}$ with $\mathsf{t} \,{\in}\, \{\mathsf{2B,A,D}\}$ and $\mathsf{s} \,{\in}\, \{\mathsf{S,W}\}$, the respective outage probability is then given as \begin{align} \label{eq:outage_sut_1}
\mathsf{P}_\mathsf{s,t}^\mathsf{o,SUT} &= \mathsf{Pr} \Big( \log_2 \left( 1\,{+}\,\rho \, |\textbf{h}_{k_\mathsf{SUT}}^{\rm H}\textbf{w}|^2 \right) \leq \overline{\mathsf{R}}_\mathsf{SUT} \Big), 
\end{align} 
where $\mathsf{\overline{R}_{SUT}}$ denotes the QoS-based target rate for any UE when scheduled for SUT. Note that the outage probability in \eqref{eq:outage_sut_1} is common to any UE in $\mathcal{S}_\mathsf{s}^\mathsf{t}$ since the particular UE with the index $k_\mathsf{SUT}$ is picked up arbitrarily. 

\subsection{Combined NOMA and Respective SUT} \label{sec:combined_noma}

As an improvement over the one-bit and two-bit NOMA schemes considered in Section~\ref{sec:one-bit-two-bit-noma}, we now propose a unified NOMA scheme, referred to as \textit{combined NOMA}, where the UE pairs are formed by following a strategy that is a combination of one-bit and two-bit NOMA schemes. This new strategy is inspired by the observation that there are certain occasions where a UE pair is possible through one-bit feedback (of either angle or distance) while no UE pair is available with two-bit NOMA (when both the feedback bits are employed). In order to leverage the superiority of NOMA over SUT in terms of spectral efficiency, the combined NOMA strategy therefore aims at forming a UE pair by following two-bit and one-bit NOMA strategies, in sequence, before switching to the respective SUT scheme. By this way, the overall transmission mechanism makes use of the spectral efficiency benefit of NOMA as much as possible through different UE pairing schemes instead of directly switching to spectrally less effective SUT schemes.

Since this new NOMA strategy assumes the availability of both the distance and angle feedback together, we adjust the respective SUT scheme to consider all the deployment possibilities. More specifically, the SUT scheme of this new framework first tries to schedule UE from the set $\mathcal{S}_\mathsf{S}^\mathsf{2B}$, and then considers $\mathcal{S}_\mathsf{S}^\mathsf{t}$ and $\mathcal{S}_\mathsf{W}^\mathsf{t}$, in sequence, if $\mathcal{S}_\mathsf{S}^\mathsf{2B}$ is an empty group, where $\mathsf{t} \,{\in}\, \{\mathsf{A,D}\}$. As a result, the SUT mechanism processes the order $\mathcal{S}_\mathsf{S}^\mathsf{2B} \rightarrow \mathcal{S}_\mathsf{S}^\mathsf{t} \rightarrow \mathcal{S}_\mathsf{W}^\mathsf{t}$ while picking up a UE to schedule in a way that whenever a non-empty group is found along this order (e.g., $\mathcal{S}_\mathsf{S}^\mathsf{2B}$), it randomly chooses a UE within that group without proceeding to the next one (e.g., $\mathcal{S}_\mathsf{S}^\mathsf{t}$ or $\mathcal{S}_\mathsf{W}^\mathsf{t}$). Note that this scheme implicitly assumes that the UE with the best channel condition is more likely to be found by employing both the distance and angle information (i.e., within $\mathcal{S}_\mathsf{S}^\mathsf{2B}$), which is consistent with the effective channel gain definition in \eqref{eq:effective_channel_gain}.  

\begin{table}[!t]
	\caption{UE Deployment Scenarios and Feasible Strategies}
	\label{table:user_deployment}
	\centering
	\begin{tabular}{|l|cccc|cc|cc|cc|}
		\cline{2-11}
		\multicolumn{1}{c}{} & \multicolumn{4}{|c}{User Region} & \multicolumn{2}{|c|}{C-NOMA} & \multicolumn{4}{|c|}{C-SUT} \\[.4ex]
		\hline 
		$\#$ & $\mathcal{S}_\mathsf{S}^\mathsf{2B}$ & $\mathcal{S}_\mathsf{W}^\mathsf{2B}$ & $\mathcal{S}_\mathsf{\overline{S}}^\mathsf{2B}$ & $\mathcal{S}_\mathsf{\overline{W}}^\mathsf{2B}$ & A & D & $\mathcal{S}_\mathsf{S}^\mathsf{A}$ & $\mathcal{S}_\mathsf{W}^\mathsf{A}$ & $\mathcal{S}_\mathsf{S}^\mathsf{D}$ & $\mathcal{S}_\mathsf{W}^\mathsf{D}$ \\[.4ex]
		\hline \hline
		1 
		& -- & -- & -- & -- 
		& -- & --  
		& -- & -- & -- & -- \\
		2 
		& -- & -- & -- & \checkmark 
		& -- & --  
		& \checkmark & -- & -- & \checkmark \\
		3 
		& -- & -- & \checkmark & -- 
		& -- & --  
		& -- & \checkmark & \checkmark & -- \\
		4 
		& -- & -- & \checkmark & \checkmark 
		& \checkmark & \checkmark  
		& \multicolumn{2}{|c|}{N/A} & \multicolumn{2}{|c|}{N/A} \\
		5 
		& -- & \checkmark & -- & -- 
		& -- & --  
		& -- & \checkmark & -- & \checkmark \\
		6 
		& -- & \checkmark & -- & \checkmark 
		& \checkmark & --  
		& \multicolumn{2}{|c|}{N/A} & -- & \checkmark \\
		7 
		& -- & \checkmark & \checkmark & -- 
		& -- & \checkmark  
		& -- & \checkmark & \multicolumn{2}{|c|}{N/A} \\
		8
		& -- & \checkmark & \checkmark & \checkmark 
		& \checkmark & \checkmark 
		& \multicolumn{2}{|c|}{N/A} & \multicolumn{2}{|c|}{N/A} \\
		9 
		& \checkmark & -- & -- & -- 
		& -- & --  
		& -- & -- & -- & -- \\
		10 
		& \checkmark & -- & -- & \checkmark 
		& -- & \checkmark  
		& -- & -- & \multicolumn{2}{|c|}{N/A} \\
		11 
		& \checkmark & -- & \checkmark & -- 
		& \checkmark & -- 
		& \multicolumn{2}{|c|}{N/A} & -- & -- \\
		12 
		& \checkmark & -- & \checkmark & \checkmark 
		& \checkmark & \checkmark 
		& \multicolumn{2}{|c|}{N/A} & \multicolumn{2}{|c|}{N/A} \\
		\hline
	\end{tabular}
\end{table}

We illustrate a subset of possible UE deployment scenarios in Table~\ref{table:user_deployment}, for which two-bit NOMA is not possible (i.e., both $\mathcal{S}_\mathsf{S}^\mathsf{2B}$ and $\mathcal{S}_\mathsf{W}^\mathsf{2B}$ are empty), over the user region of Fig.~\ref{fig:setting}. More specifically, we describe the feasibility of combined NOMA (C-NOMA) involving angle-based (A) or distance-based (D) one-bit NOMA along with the UE groups $\mathcal{S}_\mathsf{S}^\mathsf{2B}$,  $\mathcal{S}_\mathsf{W}^\mathsf{2B}$, $\mathcal{S}_\mathsf{\overline{S}}^\mathsf{2B}$, and $\mathcal{S}_\mathsf{\overline{W}}^\mathsf{2B}$ being empty (--) or nonempty ($\checkmark$). We furthermore illustrate the UE group for each deployment scenario which includes the particular UE scheduled for the respective SUT (C-SUT) scheme. As an example, two-bit NOMA is not feasible for the deployment scenario 6, where the nonempty UE groups are $\mathcal{S}_\mathsf{W}^\mathsf{2B}$ and $\mathcal{S}_\mathsf{\overline{W}}^\mathsf{2B}$, for which C-NOMA is feasible with angle-based one-bit NOMA. In addition, since C-NOMA with distance-based one-bit NOMA is not feasible, the respective C-SUT scheme schedules a UE from the group $\mathcal{S}_\mathsf{W}^\mathsf{D} \,{=}\, \mathcal{S}_\mathsf{W}^\mathsf{2B} \,{\cup}\, \mathcal{S}_\mathsf{\overline{W}}^\mathsf{2B}$ for this particular scenario.      

\section{Hybrid Sum Rate Performance} \label{sec:hybrid_rates}

In this section, we first describe the hybrid sum rates for the one-bit, two-bit, and combined NOMA strategies together with the respective SUT schemes. We then consider derivation of the outage and occurrence probabilities for each of these schemes to characterize the resulting hybrid sum rates. To this end, we also derive the analytical CDF expression for the effective channel gain associated with each particular strategy, which is required by the rate derivations.

\subsection{Hybrid Sum Rate}

We refer the overall sum rate involving each particular NOMA strategy and the associated SUT scheme as \textit{hybrid sum rate}. In the following, we first formulate hybrid sum rate for one- or two-bit NOMA (along with respective SUT schemes), and then consider the case involving combined NOMA. Assuming one-bit (with either angle or distance feedback) or two-bit NOMA, the respective sum rate can be represented using \eqref{eq:outage_noma_strong} and \eqref{eq:outage_noma_weak} as follows
\begin{align} 
\mathsf{R_{NOMA}^t} &= \sum\limits_{n=2}^{\infty} \mathsf{Pr} \left( K{=}\,n \right) \left(  \left( 1{-}\mathsf{P}_\mathsf{S,t}^\mathsf{o,NOMA} \right) \overline{\mathsf{R}}_{\mathsf{S}} + \left( 1{-}\mathsf{P}_\mathsf{W,t}^\mathsf{o,NOMA} \right) \overline{\mathsf{R}}_{\mathsf{W}}\right), \\
&= \left( 1 - {\rm e}^\mu \left( 1 + \mu \right) \right) \left(  \left( 1{-}\mathsf{P}_\mathsf{S,t}^\mathsf{o,NOMA} \right) \overline{\mathsf{R}}_{\mathsf{S}} + \left( 1{-}\mathsf{P}_\mathsf{W,t}^\mathsf{o,NOMA} \right) \overline{\mathsf{R}}_{\mathsf{W}}\right),
\label{eq:sumrate_noma}
\end{align} 
which relies on the fact that outage probability of either strong or weak UE does not depend on the total number of UEs which is Poisson distributed, and $\mathsf{t} \,{\in}\, \{\mathsf{2B,A,D}\}$. The respective rate of SUT with the scheduled UE from $\mathcal{S}_\mathsf{s}^\mathsf{t}$ is similarly given by \eqref{eq:outage_sut_1} as follows
\begin{align} \mathsf{R_{SUT}^{s,t}} &= \sum\limits_{n=1}^{\infty} \mathsf{Pr} \left( K{=}\,n \right) \left( 1{-}\mathsf{P}_\mathsf{s,t}^\mathsf{o,SUT} \right) \overline{\mathsf{R}}_\mathsf{SUT}, \\
&= \left( 1 - {\rm e}^\mu \right) \left( 1{-}\mathsf{P}_\mathsf{s,t}^\mathsf{o,SUT} \right) \overline{\mathsf{R}}_\mathsf{SUT},
\label{eq:sumrate_sut}
\end{align}
where $\mathsf{t} \,{\in}\, \{\mathsf{2B,A,D}\}$ and $\mathsf{s} \,{\in}\, \{\mathsf{S,W}\}$. Recall that the SUT scheme associated with either one-bit or two-bit NOMA searches the appropriate strong and the weak UE groups with the order $\mathcal{S}_\mathsf{S}^\mathsf{t}\rightarrow \mathcal{S}_\mathsf{W}^\mathsf{t}$, where the individual rate associated with each of these UE groups is represented by \eqref{eq:sumrate_sut}. As a result, the respective hybrid sum rate for any one-bit or two-bit strategy is given by the help of \eqref{eq:sumrate_noma} and \eqref{eq:sumrate_sut} as follows
\begin{align} \label{eq:sumrate_hybrid}
\mathsf{R_{HYB}^t} &= \mathsf{P_{NOMA}^t} \, \mathsf{R_{NOMA}^t} + \sum_\mathsf{s \in \{S,W\}} \mathsf{P_{SUT}^{s,t}} \, \mathsf{R_{SUT}^{s,t}},
\end{align} 
where $\mathsf{P_{NOMA}^t}$ and $\mathsf{P_{SUT}^{s,t}}$ stand for the probability of occurrence of appropriate one-bit or two-bit NOMA and the respective SUT scheme, respectively, with $\mathsf{t} \,{\in}\, \{\mathsf{2B,A,D}\}$. Note that the hybrid sum rate in \eqref{eq:sumrate_hybrid} can also be viewed as a weighted sum of conditional NOMA and SUT rates.

The combined NOMA strategy consists of two- and one-bit NOMA, which are considered in sequence while setting up a UE pair, and a SUT scheme searching $\mathcal{S}_\mathsf{S}^\mathsf{2B} \rightarrow \mathcal{S}_\mathsf{S}^\mathsf{t} \rightarrow \mathcal{S}_\mathsf{W}^\mathsf{t}$ to find the best UE to schedule. Whenever the overall strategy ends up with a UE pair through one-bit or two-bit NOMA strategy, the respective sum rate can still be expressed by \eqref{eq:sumrate_noma}. Similarly, the rate expression given in \eqref{eq:sumrate_sut} still applies to the SUT scheme irrespective of which particular UE group the scheduled UE belongs to. 

As a result, the hybrid rate of combined NOMA is given by \eqref{eq:sumrate_noma} and \eqref{eq:sumrate_sut} as follows
\begin{align} \label{eq:sumrate_hybrid_combined}
\mathsf{R_{C-HYB}^t} &= \underbrace{ \vphantom{\sum\nolimits_\mathsf{s \in \{S,W\}}} \mathsf{P_{NOMA}^{2B}} \, \mathsf{R_{NOMA}^{2B}} + \mathsf{P_{C-NOMA}^t} \, \mathsf{R_{NOMA}^t} }_{\text{rate for NOMA stage}} + \underbrace{\mathsf{P_{C-SUT}^{S,2B,t}} \, \mathsf{R_{SUT}^{S,2B}} + \sum\nolimits_\mathsf{s \in \{S,W\}} \mathsf{P_{C-SUT}^{s,t}} \, \mathsf{R_{SUT}^{s,t}}}_{\text{rate for SUT stage}},
\end{align} 
where $\mathsf{P_{C-NOMA}^t}$ is the occurrence probability of the respective one-bit NOMA, and $\mathsf{P_{C-SUT}^{S,2B,t}}$ and $\mathsf{P_{C-SUT}^{s,t}}$ are the occurrence probability of the SUT scheme with the scheduled UE from $\mathcal{S}_\mathsf{S}^\mathsf{2B}$ and $\mathcal{S}_\mathsf{s}^\mathsf{t}$, respectively, with $\mathsf{t} \,{\in}\, \{\mathsf{D,A}\}$. We note that the occurrence probabilities---except that for the two-bit NOMA---associated with combined NOMA scheme are different from the individual one-bit and two-bit NOMA occurrence probabilities appearing in \eqref{eq:sumrate_hybrid} as they take into account whether any of the previous multiplexing schemes happen or not.

\subsection{Distribution of Effective Channel Gain}

The distribution of the effective channel gain of a UE strongly depends on the user group that the UE belongs to. In the following, we give the CDF of the effective channel gain which enables the computation outage probabilities for any type of NOMA and SUT scheme under consideration.

\begin{theorem}\label{theorem:cdf_ecg}
The CDF of the effective channel gain for the $k$-th UE with $k \,{\in}\, \mathcal{S}_\mathsf{s}^\mathsf{t}$, where $\mathsf{s} \,{\in}\, \{\mathsf{W,S}\}$ denotes the UE being weak or strong, respectively, $\mathsf{t} \,{\in}\, \{\mathsf{2B, A,D}\}$ stands for the feedback type being two-bit, one-bit angle, and one-bit distance, respectively, is given as follows 
\begin{align}\label{eq:cdf_ecg}
    \!\!\mathsf{F}_\mathsf{s}^{\,\mathsf{t}}(x) \,{=}\, \frac{1}{\xi_\mathsf{s}^\mathsf{t}} \int\limits_{\mathcal{R}_d^\mathsf{s,t}} \int\limits_{\mathcal{R}_\theta^\mathsf{s,t}} \!\!\left( 1 - \exp \left\lbrace {-}\frac{\mathsf{PL}(r)}{\mathsf{F}_M(\overline{\theta},\theta )}\frac{x}{\sigma^2} \right\rbrace \right) r \dd r  \dd \theta ,
\end{align}
where $\xi_\mathsf{S}^\mathsf{2B} \,{=}\,\theta_\mathsf{th} \!\left( d_\mathsf{th}^2 \,{-}\, d_\mathsf{min}^2\right)$, $\xi_\mathsf{W}^\mathsf{2B} \,{=}\, \left( \frac{\Delta}{2} \,{-}\, \theta_\mathsf{th}\right) \!\left( d_\mathsf{max}^2 {-}\, d_\mathsf{th}^2\right)$, $\xi_\mathsf{S}^\mathsf{D} \,{=}\, \frac{\Delta}{2} \!\left( d_\mathsf{th}^2 \,{-}\, d_\mathsf{min}^2\right)$, $\xi_\mathsf{W}^\mathsf{D} \,{=}\, \frac{\Delta}{2} \!\left( d_\mathsf{max}^2 {-}\, d_\mathsf{th}^2\right)$, $\xi_\mathsf{S}^\mathsf{A} \,{=}\,\theta_\mathsf{th} \!\left( d_\mathsf{max}^2 \,{-}\, d_\mathsf{min}^2\right)$, and $\xi_\mathsf{W}^\mathsf{A} \,{=}\, \left( \frac{\Delta}{2} \,{-}\, \theta_\mathsf{th}\right) \!\left( d_\mathsf{max}^2 {-}\, d_\mathsf{min}^2\right)$ are the normalization coefficients, $\mathcal{R}_d^\mathsf{W,2B} = \mathcal{R}_d^\mathsf{W,D} \,{=}\, [d_\mathsf{th},d_\mathsf{max}]$, $\mathcal{R}_d^\mathsf{S,2B} \,{=}\, \mathcal{R}_d^\mathsf{S,D} \,{=}\, [d_\mathsf{min},d_\mathsf{th}]$, and $\mathcal{R}_d^\mathsf{W,A} \,{=}\, \mathcal{R}_d^\mathsf{S,A} \,{=}\, [d_\mathsf{min},d_\mathsf{max}]$ are the regions of integration for the distance domain, and $\mathcal{R}_\theta^\mathsf{W,D} \,{=}\, \mathcal{R}_\theta^\mathsf{S,D} \,{=}\, [\overline{\theta} - \frac{\Delta}{2},\overline{\theta} + \frac{\Delta}{2}]$, $\mathcal{R}_\theta^\mathsf{S,2B} \,{=}\, \mathcal{R}_\theta^\mathsf{S,A} \,{=}\, [\overline{\theta} - \theta_\mathsf{th},\overline{\theta} + \theta_\mathsf{th}]$, and $\mathcal{R}_\theta^\mathsf{W,2B} \,{=}\, \mathcal{R}_\theta^\mathsf{W,A} \,{=}\, \mathcal{R}_\theta^\mathsf{S,D} {\setminus} \mathcal{R}_\theta^\mathsf{S,2B}$ are the regions of integration for the angular domain.
\end{theorem}
\begin{IEEEproof}
See Appendix~\ref{appendix:cdf_ecg}.
\end{IEEEproof}

Note that the integration in \eqref{eq:cdf_ecg} is overly complicated for any feedback scheme to obtain a closed-form expression due to the Fej\'er Kernel function, but is computable using the numerical integration methods available in the literature.  

\subsection{Outage Probabilities}
The outage probability for the strong UE of any NOMA scheme is given in terms of the CDF of the effective channel gain in \eqref{eq:cdf_ecg} and by using the definition in \eqref{eq:outage_noma_strong} as follows 
\begin{align} 
\mathsf{P}_\mathsf{S,t}^\mathsf{o,NOMA} &= 1 - \mathsf{Pr} \left( |\textbf{h}_{k_\mathsf{S}}^{\rm H}\textbf{w}|^2 > \frac{\epsilon_{\mathsf{W}}/\rho}{ \beta_\mathsf{W}^2-\epsilon_{\mathsf{W}}\beta_\mathsf{S}^2 }, |\textbf{h}_{k_\mathsf{S}}^{\rm H}\textbf{w}|^2 > \frac{\epsilon_{\mathsf{S}}/\rho}{ \beta_\mathsf{S}^2-\epsilon_{\mathsf{S}}\beta_\mathsf{W}^2 }  \right) \label{eq:outage_noma_strong_2}\\
&= \mathsf{F}_\mathsf{S}^\mathsf{t} \left( \frac{\mathsf{max} \left(\eta_\mathsf{W},\eta_\mathsf{S}\right)}{\rho} \right), \label{eq:outage_noma_strong_3}
\end{align} 
where $\eta_\mathsf{S} \,{=}\, \epsilon_{\mathsf{S}} {/} \left( \beta_\mathsf{S}^2{-}\epsilon_{\mathsf{S}}\beta_\mathsf{W}^2 \right)$, $\eta_\mathsf{W} \,{=}\, \epsilon_{\mathsf{W}} {/} \left( \beta_\mathsf{W}^2{-}\epsilon_{\mathsf{W}}\beta_\mathsf{S}^2 \right)$, and $\beta_\mathsf{S}$ ($\beta_\mathsf{W}$) is the power allocation coefficient for the strong (weak) UE. Similarly, outage probability for the weak UE is given by using \eqref{eq:outage_noma_weak} as follows 
\begin{align} 
\mathsf{P}_\mathsf{W,t}^\mathsf{o,NOMA} &= \mathsf{Pr} \left( |\textbf{h}_{k_\mathsf{W}}^{\rm H}\textbf{w}|^2 \leq \frac{\epsilon_{\mathsf{W}}/\rho}{ \beta_\mathsf{W}^2-\epsilon_{\mathsf{W}}\beta_\mathsf{S}^2 } \right) = \mathsf{F}_\mathsf{W}^\mathsf{t} \left( \frac{\eta_\mathsf{W}}{\rho} \right), \label{eq:outage_noma_weak_2}
\end{align} 

The outage probability for SUT transmission defined in \eqref{eq:outage_sut_1} is further elaborated as follows
\begin{align} \label{eq:outage_sut_2}
\mathsf{P}_\mathsf{s,t}^\mathsf{o,SUT} &= \mathsf{Pr} \left(  |\textbf{h}_{k_\mathsf{SUT}}^{\rm H}\textbf{w}|^2 \leq \frac{\epsilon_{\mathsf{SUT}}}{\rho} \right) =  \mathsf{F}_\mathsf{s}^\mathsf{t} \left( \frac{\epsilon_{\mathsf{SUT}}}{\rho} \right), 
\end{align} 
where $\epsilon_{\mathsf{SUT}}\,{=}\,2^{\overline{\mathsf{R}}_\mathsf{SUT}}\,{-}\,1$. As a result, the outage probabilities given by \eqref{eq:outage_noma_strong_3}, \eqref{eq:outage_noma_weak_2}, and \eqref{eq:outage_sut_2} can all be computed by the CDF of \eqref{eq:cdf_ecg}. Note also that employing \eqref{eq:outage_noma_weak_2} and \eqref{eq:outage_noma_strong_3} in \eqref{eq:sumrate_noma}, and \eqref{eq:outage_sut_2} in \eqref{eq:sumrate_sut}, one can obtain the individual sum rates for NOMA and SUT transmission, respectively, which yield the overall hybrid sum rates in \eqref{eq:sumrate_hybrid} and \eqref{eq:sumrate_hybrid_combined}. 

\subsection{Occurrence Probabilities}
We finally consider the occurrence probability of NOMA and SUT cases in order to weight the respective conditional data rates to obtain the hybrid sum rates defined in \eqref{eq:sumrate_hybrid} and \eqref{eq:sumrate_hybrid_combined}. To this end, we first consider one-bit and two-bit NOMA, and then switch to combined NOMA. For the ease of presentation, we consider SUT occurrence probabilities before those of NOMA.  

\begin{theorem} \label{theorem:event_prob_sut}
The occurrence probability of SUT associated with two-bit NOMA where the scheduled UE is picked up from $\mathcal{S}_\mathsf{s}^\mathsf{2B}$ with $\mathsf{s} \,{\in}\, \{\mathsf{S,W}\}$ is given as
\begin{align}\label{eq:prob_sut_twobit}
    \mathsf{P}_\mathsf{SUT}^\mathsf{s,2B} = {\rm e}^{{-}\mu \left( 1 - p_\mathsf{s,2B} \right)} - {\rm e}^{{-}\mu \left( 1 - p_\mathsf{\overline{2B}} \right)} ,
\end{align}
where $p_\mathsf{S,2B} \,{=}\, p_\theta p_d$ with $p_\theta \,{=}\, \mathsf{Pr} \left( \theta_k \,{\leq}\, \theta_\mathsf{th} \right)\,{=}\, \frac{2 \theta_\mathsf{th}}{\Delta}$ and $p_d \,{=}\, \mathsf{Pr} \left( d_k \,{\leq}\, d_\mathsf{th} \right) \,{=}\, (d_\mathsf{th}^2 {-} d_\mathsf{min}^2)(d_\mathsf{max}^2{-}d_\mathsf{min}^2)$, $p_\mathsf{W,2B} \,{=}\, (1{-}p_\theta)(1{-}p_d)$, and $p_\mathsf{\overline{2B}} \,{=}\, 1 \,{-}\, p_\mathsf{S,2B} \,{-}\, p_\mathsf{W,2B}$. Similarly, the desired occurrence probability of SUT associated with one-bit NOMA for which the scheduled UE belongs to $\mathcal{S}_\mathsf{s}^\mathsf{t}$ with $\mathsf{t} \,{\in}\, \{\mathsf{A,D}\}$ (i.e., applies to both the angle- and distance-only feedback) and $\mathsf{s} \,{\in}\, \{\mathsf{S,W}\}$ is 
\begin{align}\label{eq:prob_sut_one}
    \mathsf{P}_\mathsf{SUT}^\mathsf{s,t} &= {\rm e}^{{-}\mu \left( 1 - p_\mathsf{s,t} \right)} - {\rm e}^{{-}\mu} ,
\end{align}
where $p_\mathsf{S,A} \,{=}\, p_\theta$, $p_\mathsf{W,A} \,{=}\, 1 \,{-}\, p_\theta$, $p_\mathsf{S,D} \,{=}\, p_d$, and $p_\mathsf{W,D} \,{=}\, 1 \,{-}\, p_d$.
\end{theorem}
\begin{IEEEproof}
See Appendix~\ref{appendix:event_prob_sut}.
\end{IEEEproof}

\begin{theorem} \label{theorem:event_prob_noma}
The occurrence probability of two-bit NOMA transmission is given by
\begin{align} \label{eq:prob_noma_twobit}
    \mathsf{P}_\mathsf{NOMA}^\mathsf{2B} &= 1 + {\rm e}^{{-}\mu \left( p_\mathsf{S,2B} + p_\mathsf{W,2B} \right)} - {\rm e}^{{-}\mu p_\mathsf{S,2B}} - {\rm e}^{{-}\mu p_\mathsf{W,2B}},
\end{align}
and that of one-bit NOMA with either angle- or distance-only feedback is given by
\begin{align}\label{eq:prob_noma_onebit}
    \mathsf{P}_\mathsf{NOMA}^\mathsf{t} &= 1 + {\rm e}^{{-}\mu} - {\rm e}^{{-}\mu p_\mathsf{S,t}} - {\rm e}^{{-}\mu p_\mathsf{W,t}},
\end{align}
where $\mathsf{t} \,{\in}\, \{\mathsf{A,D}\}$, and $p_\mathsf{s,t}$'s and $p_\mathsf{s,2B}$'s for $\mathsf{s} \,{\in}\, \{\mathsf{S,W}\}$ are all defined in Theorem~\ref{theorem:event_prob_sut}.
\end{theorem}
\begin{IEEEproof}
See Appendix~\ref{appendix:event_prob_noma}.
\end{IEEEproof}

\begin{theorem} \label{theorem:prob_sut_comb}
The probability for the combined NOMA transmitter of scheduling a UE from the group $\mathcal{S}_\mathsf{S}^\mathsf{2B}$ during SUT is given as
\begin{align}
    \mathsf{P}_{\mathsf{C-SUT}}^\mathsf{S,2B,t} = {\rm e}^{{-}\mu\left(1-p_\mathsf{S,t}\right)} - {\rm e}^{{-}\mu\left(1-p_\mathsf{S,t} + p_\theta p_d \right)},
\label{eq:prob_sut_comb_two}
\end{align}
where $\mathsf{t} \,{\in}\, \{\mathsf{A,D}\}$, and $p_\mathsf{S,t}$'s, $p_\theta$, and $p_d$ are all defined in Theorem~\ref{theorem:event_prob_sut}.
Note that \eqref{eq:prob_sut_comb_two} applies to any feedback scheme as a part of overall combine NOMA strategy. Similarly, the SUT occurrence probabilities for a UE belonging to $\mathcal{S}_\mathsf{S}^\mathsf{t}$ and $\mathcal{S}_\mathsf{W}^\mathsf{t}$ are given, respectively, as follows
\begin{align}
    \mathsf{P}_\mathsf{C-SUT}^\mathsf{S,t} &=
    {\rm e}^{{-}\mu\left(1 - p_\mathsf{S,t} + p_\theta p_d \right)} - {\rm e}^{{-}\mu},
    \label{eq:prob_sut_comb_strong} \\
    \mathsf{P}_\mathsf{C-SUT}^\mathsf{W,t} &=
    {\rm e}^{{-}\mu p_\mathsf{S,t}} - {\rm e}^{{-}\mu}.
    \label{eq:prob_sut_comb_weak}
\end{align}
\end{theorem}
\begin{IEEEproof}
See Appendix~\ref{appendix:prob_sut_comb}.
\end{IEEEproof}

Note that the term $\mathsf{P_{C-NOMA}^t}$ representing the occurrence probability of one-bit NOMA in a combined NOMA strategy can be found by considering all the possible transmission situations, for which the respective occurrence probabilities are supposed to add up to $1$. 
We can therefore formulate $\mathsf{P_{C-NOMA}^t}$ for $\mathsf{t} \,{\in}\, \{\mathsf{A,D}\}$ by \eqref{eq:prob_noma_twobit}, \eqref{eq:prob_sut_comb_two}, \eqref{eq:prob_sut_comb_strong}, and \eqref{eq:prob_sut_comb_weak} as follows  
\begin{align} 
\mathsf{P_{C-NOMA}^t} &= 1- \mathsf{P_{NOMA}^{2B}} - \mathsf{P_{C-SUT}^{S,2B,t}} -  \mathsf{P_{C-SUT}^{S,t}} -  \mathsf{P_{C-SUT}^{W,t}} - \mathsf{P_{NO}} \label{eq:one_bit_combined_noma_1} \\
&= {\rm e}^{{-}\mu} - {\rm e}^{{-}\mu \left( 1 - p_\mathsf{S,t} \right)} - {\rm e}^{{-}\mu p_\mathsf{S,t} } - {\rm e}^{{-}\mu \left( p_\mathsf{S,2B} + p_\mathsf{W,2B} \right) } + {\rm e}^{{-}\mu p_\mathsf{S,2B} } + {\rm e}^{{-}\mu p_\mathsf{W,2B} }, \label{eq:one_bit_combined_noma_2}
\end{align}
where $p_\mathsf{S,t}$, $p_\mathsf{S,2B}$ and $p_\mathsf{W,2B}$ are given in Theorem~\ref{theorem:event_prob_sut}. Note that $\mathsf{P_{NO}} \,{=}\, {\rm e}^{{-}\mu}$ in \eqref{eq:one_bit_combined_noma_1} is the probability of no transmission, which occurs only when $K \,{=}\, 0$ since the combined NOMA strategy always ends up with either individual NOMA or SUT schemes for $K \,{\geq}\, 1$.

\section{Numerical Results} \label{sec:results}

\begin{table}[!t]
	\caption{Simulation Parameters}
	\label{tab:simulation_parameters}
	\centering
	\begin{tabular}{lc}
		\hline
		Parameter & Value \\
		\hline\hline
		UE density $(\lambda)$ & $0.01\,\text{m}^{{-}2}$ \\
		Antenna array size $(M)$ & $64$ \\
		Path-loss exponent $(\gamma)$ & $2$ \\
        Path gain variance ($\sigma^2$) & $1$\\
		Power allocation coefficients $(\beta_{\mathsf{W}},\beta_{\mathsf{S}})$ & $(0.6,0.4)$ \\
		Target rate of weak NOMA UE ($\overline{\mathsf{R}}_{\mathsf{W}}$) & $1\,\text{bps/Hz}$  \\
		Target rate of strong NOMA UE ($\overline{\mathsf{R}}_{\mathsf{S}}$) & $8\,\text{bps/Hz}$  \\
		Target rate of SUT UE ($\overline{\mathsf{R}}_{\mathsf{SUT}}$) & $8\,\text{bps/Hz}$  \\
		Center angle ($\Delta$) & $15^\circ$\\
		Inner radius ($d_\mathsf{min}$) & $0\,\text{m}$\\
		Outer radius ($d_\mathsf{max}$) & $45\,\text{m}$\\
        Precoder AoD $(\overline{\theta})$ & $0^\circ$\\
        Angle threshold coefficient ($c_\theta$) & $0.1$\\
        Distance threshold coefficient ($c_d$) & $0.2$\\
		\hline
	\end{tabular}
\end{table}

In this section, we present numerical results based on extensive Monte Carlo simulations, to evaluate the performance of the proposed NOMA strategies rigorously. Considering the mmWave propagation characteristics \cite{Rappaport2017OveMil}, the simulation parameters are listed in Table~\ref{tab:simulation_parameters} unless otherwise stated. The threshold values $d_\mathsf{th}$ and $\theta_\mathsf{th}$ are obtained using the coefficients $c_d \,{\in}\, [0,1]$ and $c_\theta \,{\in}\, [0,1]$, respectively, through the equations $d_\mathsf{th} \,{=}\, d_\mathsf{min} \,{+}\, c_d \left( d_\mathsf{max} {-}\, d_\mathsf{min} \right)$ and $\theta_\mathsf{th} \,{=}\, c_\theta \frac{\Delta}{2}$. 

\begin{figure}[!t]
\centering
\includegraphics[width=0.7\textwidth]{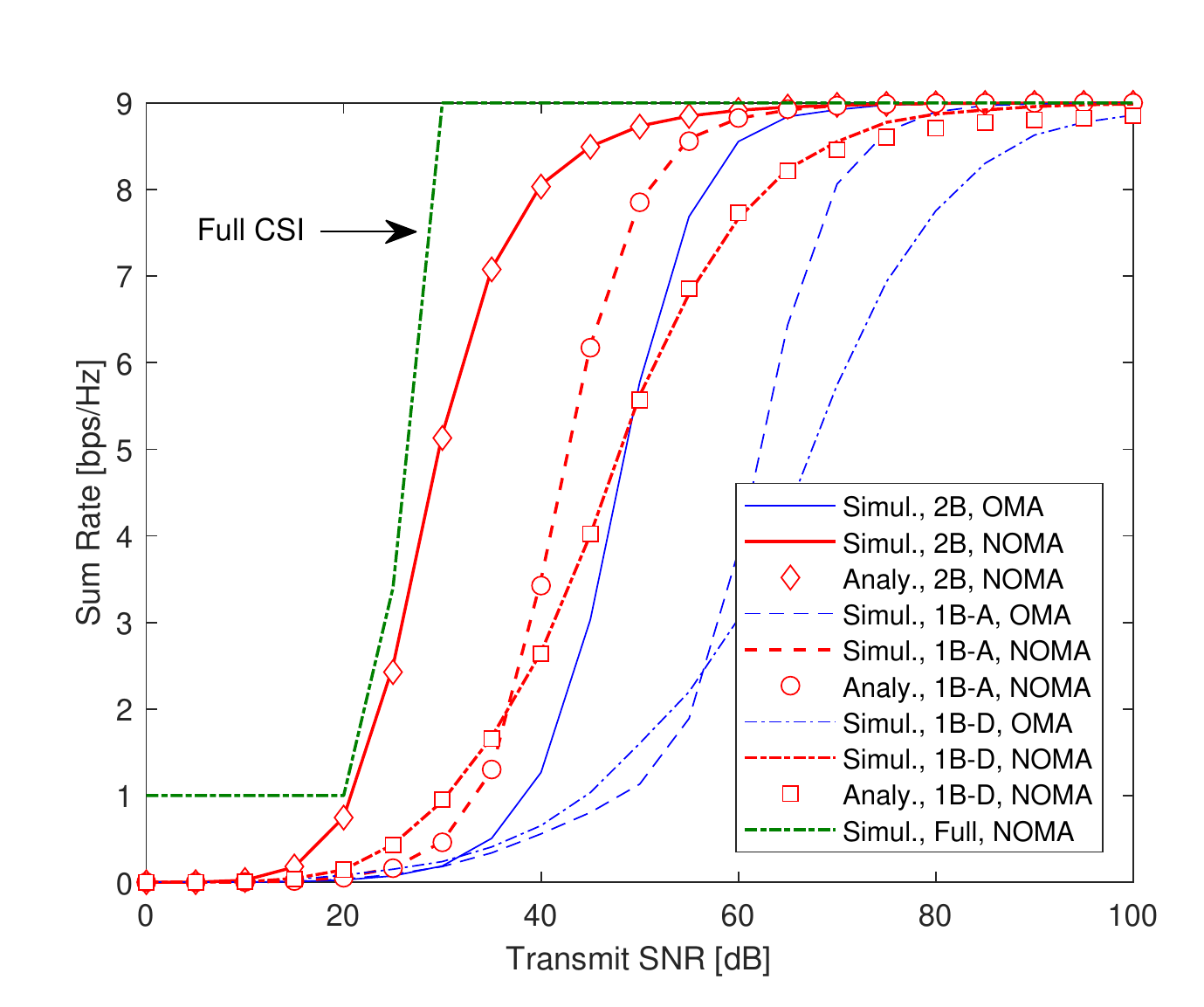}
\vspace{0.0in}
\caption{Sum rate for angle-only (1B-A) and distance-only (1B-D) one-bit NOMA, two-bit NOMA (2B), full-CSI NOMA, and OMA with $d_\mathsf{max} \,{=}\, 45\,\text{m}$ and $\Delta \,{=}\, 15^\circ$.}
\label{fig:rate_dmax45_beam15}
\vspace{-0.2in}
\end{figure}

In Fig.~\ref{fig:rate_dmax45_beam15}, we depict the sum-rate performance of one-bit and two-bit NOMA strategies along with varying transmit SNR. For comparison purposes, we also include the performance of 1) the OMA scheme for the UEs chosen for each NOMA strategy, and 2) the NOMA scheme with full channel state information (CSI), i.e., without any quantization, where the respective rates are calculated following \cite{Ding2017RanBea}. In particular, we assume that the strong and weak UEs are chosen for the full-CSI NOMA scheme as the ones with the indices $1$ and $10$, respectively, in $\mathcal{N}_\mathsf{U}$. We observe that the analytical and simulation results associated with the NOMA strategies show a very nice match, and that NOMA is significantly superior to OMA. We also observe that two-bit NOMA employing both the angle and distance feedback bits jointly is much better than one-bit NOMA strategies (with the distance or angle feedback only), and is very close to the full-CSI feedback having no quantization at all. Furthermore, one-bit NOMA with angle-only feedback yields a better performance than that of distance-only feedback for transmit SNR larger than $35\,\text{dB}$. This observation underscores the fact that although angle-only one-bit feedback is more powerful than distance-only one-bit feedback in describing UE channel quality for this particular setting, the overall NOMA performance is enhanced even further when both feedback bits are employed through the two-bit NOMA strategy. Note that NOMA transmission is always feasible for this particular setting under any feedback strategy, which implies that  $\mathsf{R_{SUT}^{s,t}} \,{=}\, 0$ and $\mathsf{R_{HYB}^t} \,{=}\, \mathsf{R_{NOMA}^t}$ in \eqref{eq:sumrate_hybrid}, and, hence, there is no need to resort to combined NOMA.  

\begin{figure}[!t]
\centering
\includegraphics[width=0.75\textwidth]{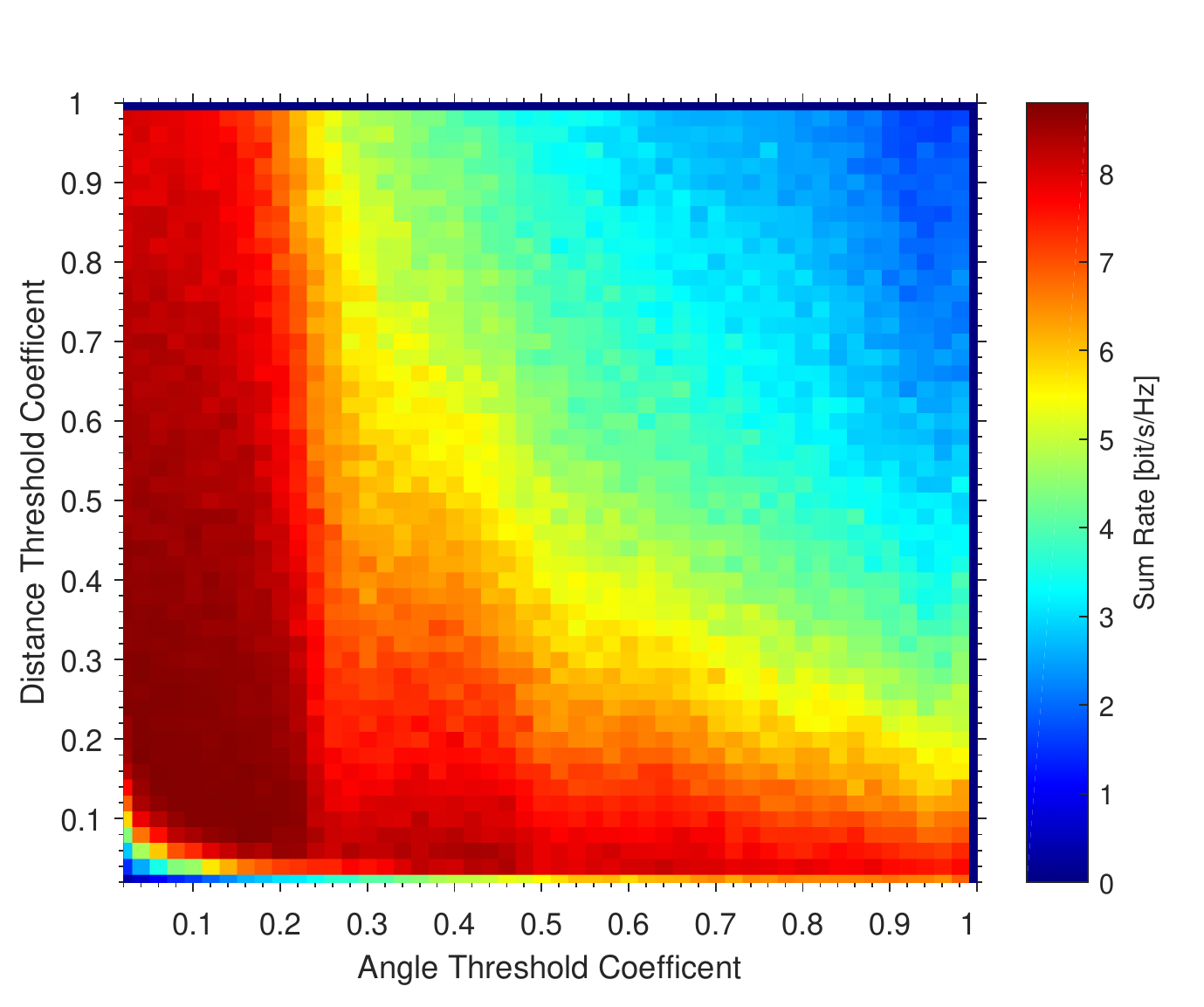}
\vspace{0.0in}
\caption{Sum rate for two-bit NOMA against varying angle and distance threshold coefficients at $50\,\text{dB}$ transmit SNR with $d_\mathsf{max} \,{=}\, 45\,\text{m}$ and $\Delta \,{=}\, 15^\circ$.}
\label{fig:twobit_optimal_threshold}
\vspace{-0.15in}
\end{figure}

We depict the sum-rate performance of two-bit NOMA in Fig.~\ref{fig:twobit_optimal_threshold} along with varying angle and distance threshold coefficients (i.e., $c_\theta$ and $c_d$, respectively). We observe that while both the coefficients should be small enough to have sufficiently stronger UEs in  $\mathcal{S}_\mathsf{S}^\mathsf{2B}$ of \eqref{eq:set_strong_2bit}, the coefficients should not be too small in order not to end up with $\mathcal{S}_\mathsf{S}^\mathsf{2B}$ having no UE at all. Note that whenever $\mathcal{S}_\mathsf{S}^\mathsf{2B}$ turns out to be an empty set, NOMA becomes unfeasible and the transmission mechanism schedules a single UE (i.e, SUT scheme), which in turn degrades sum-rate performance. The sum-rate results in Fig.~\ref{fig:twobit_optimal_threshold} show that the optimal performance can be obtained by choosing the threshold coefficients from a specific interval of values, which are different for the angel and distance coefficients. We accordingly pick up $c_\theta \,{=}\, 0.1$ and $c_d \,{=}\, 0.2$, as listed in Table~\ref{tab:simulation_parameters}, to obtain the best performance for the rest of the simulations. 

\begin{figure}[!t]
\centering
\includegraphics[width=0.7\textwidth]{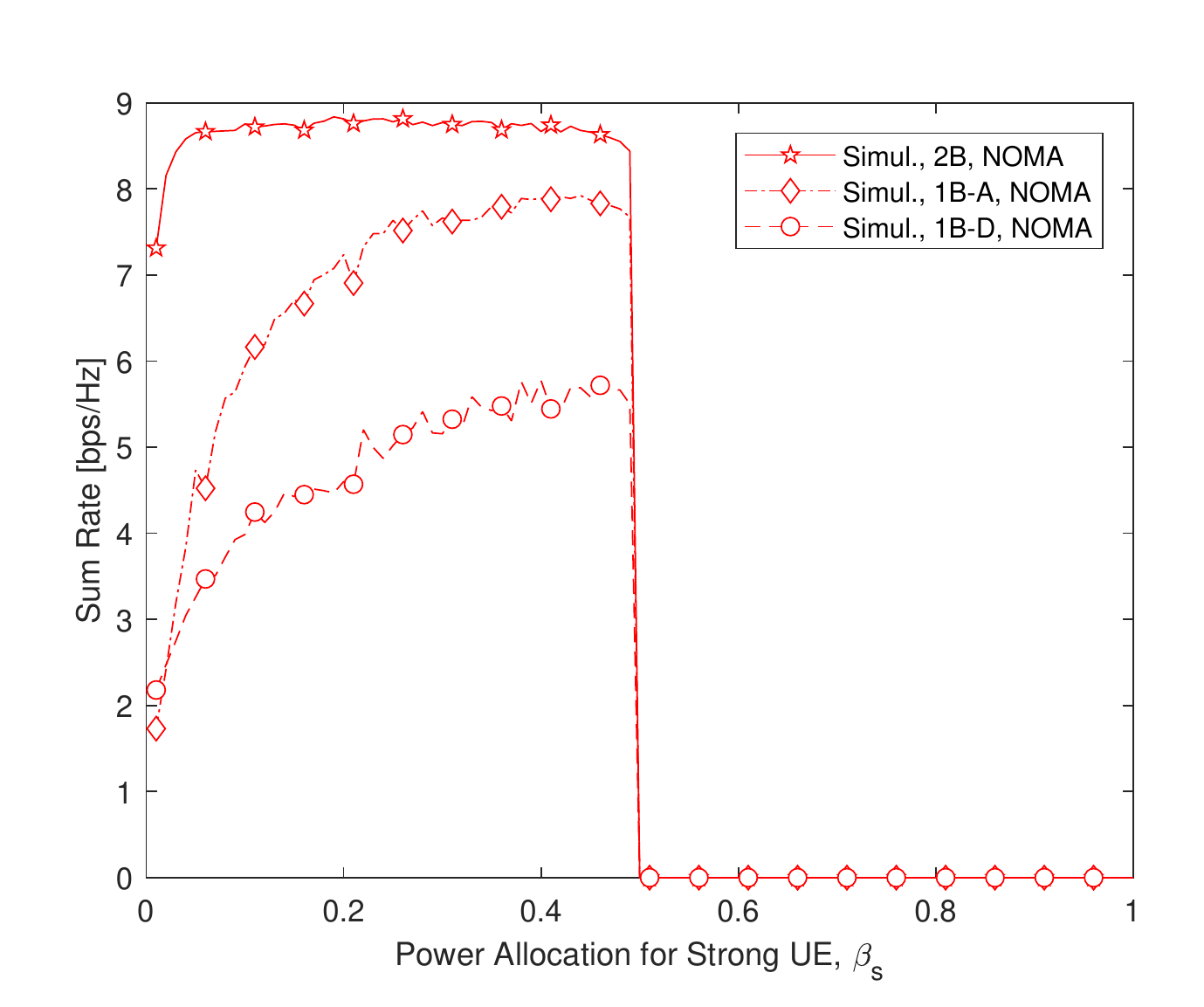}
\vspace{0.0in}
\caption{Sum rate for angle-based (1B-A) and distance-based (1B-D) one-bit NOMA, and two-bit NOMA (2B) against power allocation coefficient of strong UE ($\beta_\mathsf{S}$) at $50\,\text{dB}$ transmit SNR with $d_\mathsf{max} \,{=}\, 45\,\text{m}$ and $\Delta \,{=}\, 15^\circ$.}
\label{fig:twobit_optimal_power}
\vspace{-0.15in}
\end{figure}

In Fig.~\ref{fig:twobit_optimal_power}, we demonstrate the sum-rate performance of one-bit and two-bit NOMA strategies along with varying power allocation coefficient of strong UE. We observe that the performance of two-bit NOMA is very robust against varying power allocation while that of one-bit NOMA with either feedback gets maximized around $0.4$. Note that $\beta_\mathsf{S} \,{\geq}\, 0.5$ is not a reasonable choice since strong UE should be allocated less power as compared to that of weak UE (i.e., $\beta_\mathsf{W} \,{>}\, \beta_\mathsf{S}$). As a result, we choose $\beta_\mathsf{S} \,{=}\, 0.4$, as listed in Table~\ref{tab:simulation_parameters}, for the best performance.  

\begin{figure}[!t]
\centering
\subfloat[One-Bit and Two-Bit NOMA] {\includegraphics[width=0.7\textwidth]{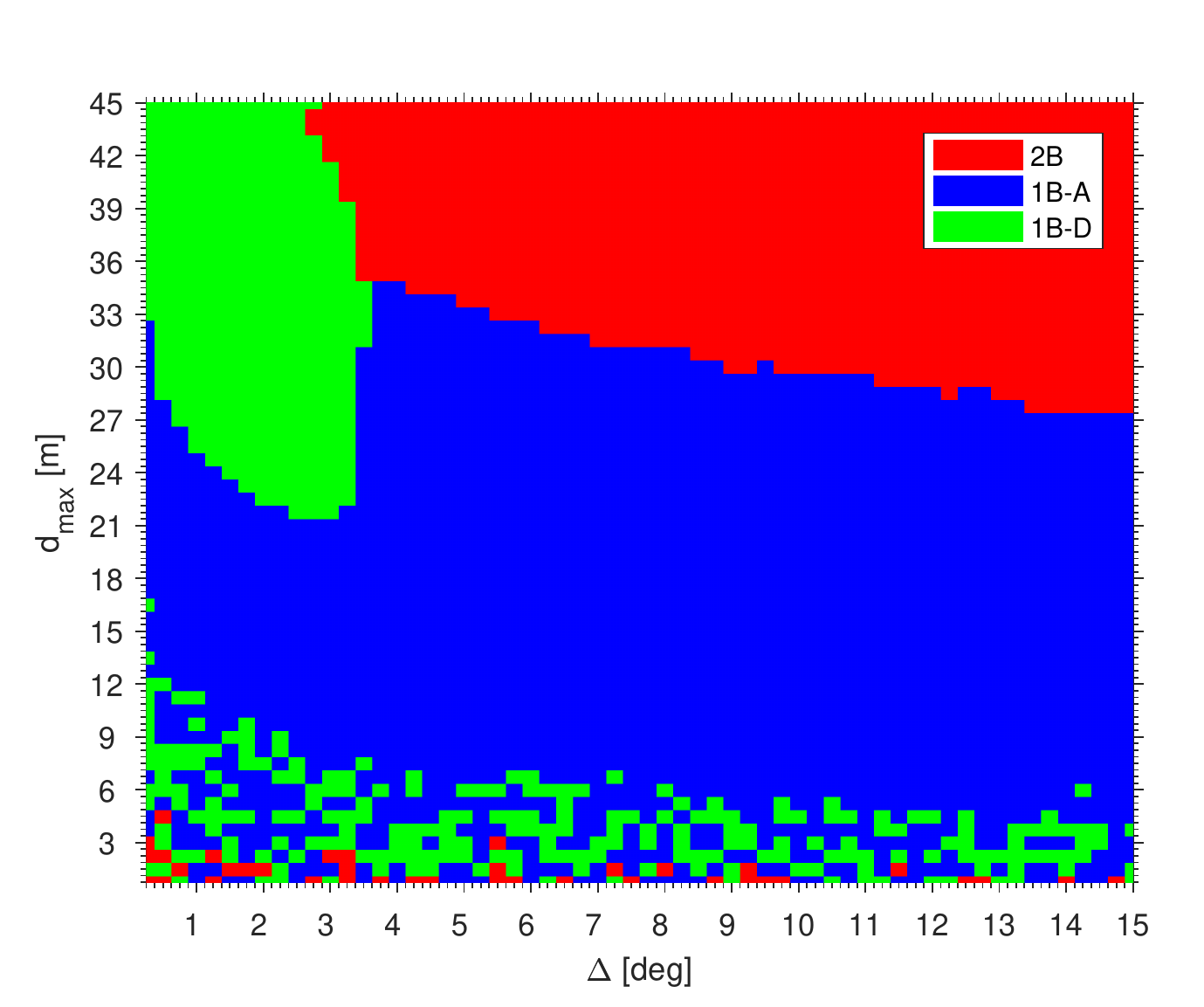}
\label{fig:geometry_noma_bw15_dmax45}}\\
\vspace{-0.25in}
\subfloat[One-Bit, Two-Bit, and Combined NOMA] {\includegraphics[width=0.7\textwidth]{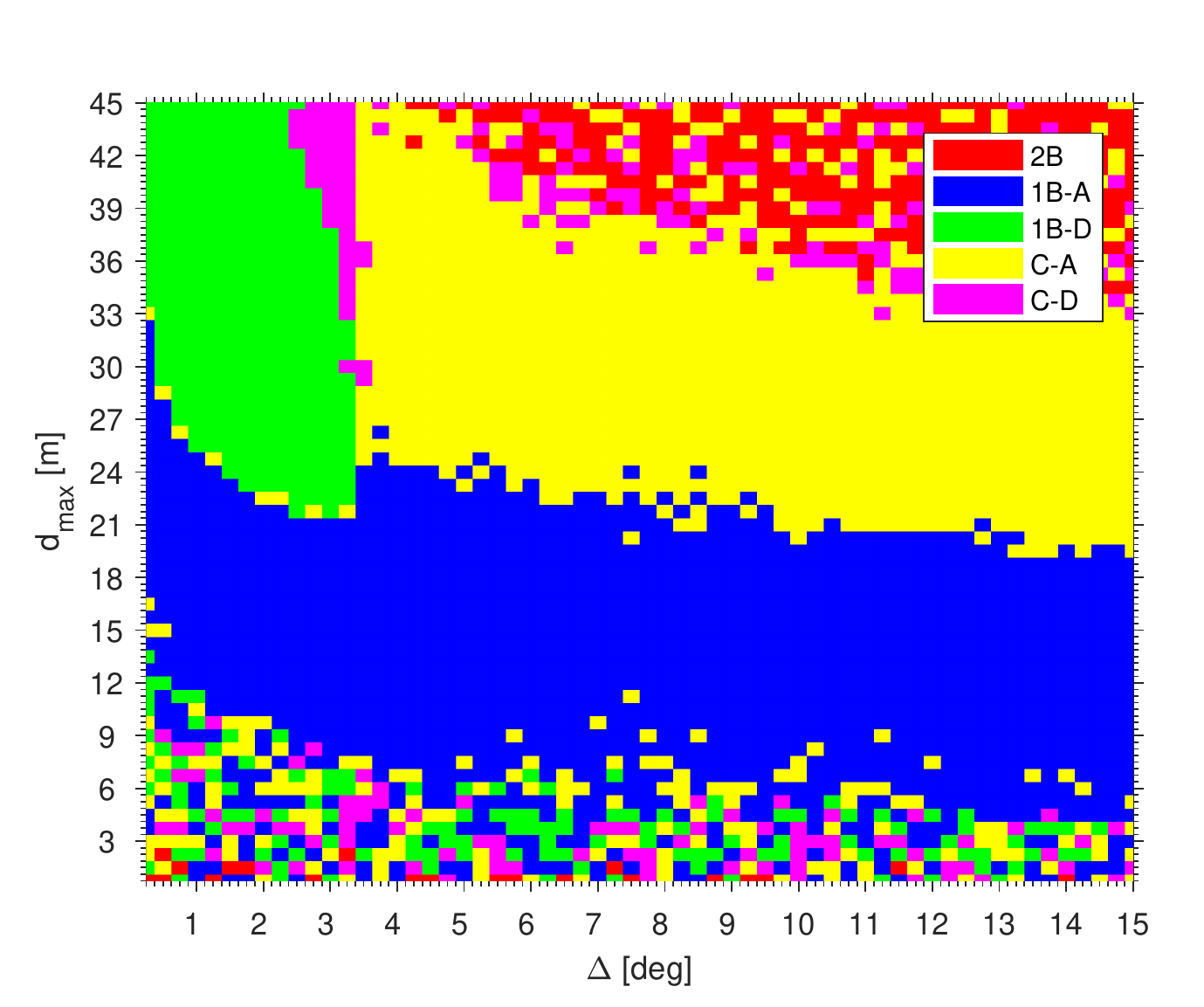}
\label{fig:geometry_cnoma_bw15_dmax45}}
\caption{Hybrid sum rates for angle-only (1B-A) and distance-only (1B-D) one-bit NOMA, angle-only (C-A) and distance-only (C-D) combined NOMA, and two-bit NOMA (2B) for varying $d_\mathsf{max}$ and $\Delta$ at $50\,\text{dB}$ transmit SNR.}
\label{fig:geometry_bw15_dmax45}
\vspace{-0.15in}
\end{figure}

We consider hybrid sum-rate performance of one-bit, two-bit, and combined NOMA strategies in Fig.~\ref{fig:geometry_bw15_dmax45} for varying size of the user region, which is controlled through the maximum outer radius $d_\mathsf{max}$ and center angle $\Delta$, as illustrated by Fig.~\ref{fig:setting}. The optimal strategy for one-bit and two-bit NOMA is indicated in Fig.~\ref{fig:geometry_bw15_dmax45}\subref{fig:geometry_noma_bw15_dmax45} while combined NOMA is involved in the assessment in Fig.~\ref{fig:geometry_bw15_dmax45}\subref{fig:geometry_cnoma_bw15_dmax45}. We observe in Fig.~\ref{fig:geometry_bw15_dmax45}\subref{fig:geometry_noma_bw15_dmax45} that two-bit NOMA is superior to both angle-only and distance-only one-bit NOMA roughly for $d_\mathsf{max} \,{\geq}\, 30\,\text{m}$ and $\Delta \,{\geq}\, 4^\circ$, for which UEs become more likely to be distinguishable when both the angle and distance information are employed. We also observe that distance-only one-bit NOMA is superior to angle-only one-bit NOMA when the user region is sufficiently large in the distance domain ($d_\mathsf{max} \,{\geq}\, 21\,\text{m}$) but not in the angle domain ($\Delta \,{\leq}\, 4^\circ$). In contrast, angle-only one-bit NOMA turns out to be the optimal strategy when UEs are deployed over a wide (angle-wise) or relatively short (distance-wise) user region (i.e., either $d_\mathsf{max} \,{<}\, 30\,\text{m}$ and $\Delta \,{\geq}\, 4^\circ$, or $d_\mathsf{max} \,{<}\, 21\,\text{m}$). 

As a result, we conclude that using one more feedback bit representing either distance or angle domain is useful for NOMA transmission only if the UEs are distinguishable in that domain (i.e., associated with the additional bit of information). In addition, if both distance and angle domain make UEs sufficiently distinctive, two-bit NOMA involving both feedback bits outperforms angle-only and distance-only one-bit feedback schemes. Note that as the user region gets shorter in radius (e.g., $d_\mathsf{max} \,{\leq}\, 9\,\text{m}$), NOMA becomes unfeasible and the transmission scheme switches to SUT. This, in turn, results in one-bit and two-bit NOMA schemes having very similar hybrid sum rates, which is the reason for the optimal strategy to change rapidly over the adjacent $(d_\mathsf{max},\Delta)$ pairs (i.e., optimal strategy is actually SUT, and effectively not changing).

\begin{figure}[!t]
\centering
\includegraphics[width=0.7\textwidth]{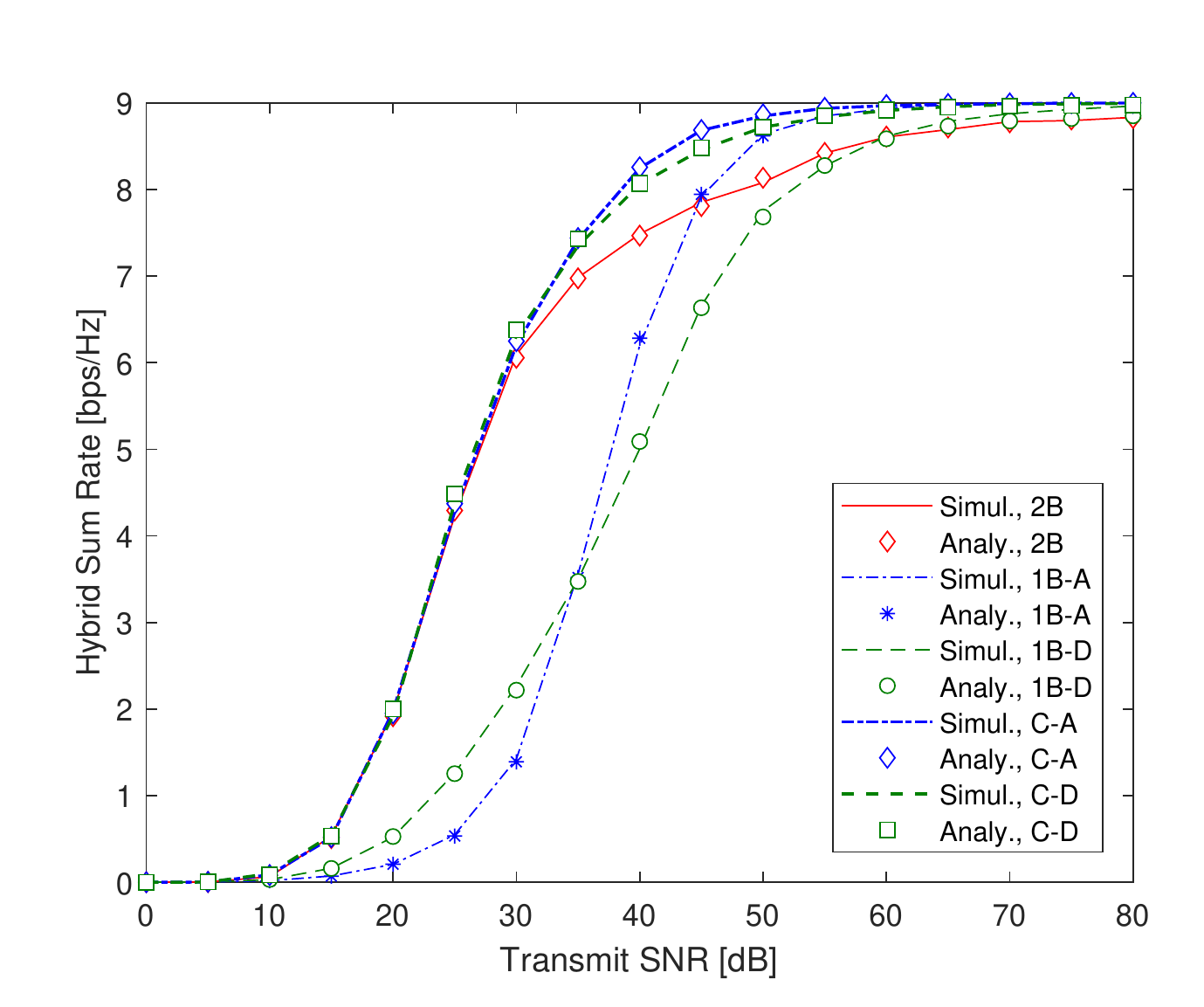}
\vspace{0.0in}
\caption{Hybrid sum rates for angle-only (1B-A) and distance-only (1B-D) one-bit NOMA, angle-only (C-A) and distance-only (C-D) combined NOMA, and two-bit NOMA (2B) for $d_\mathsf{max} \,{=}\, 27\,\text{m}$ and $\Delta \,{=}\, 10^\circ$.}
\label{fig:rate_dmax27_beam10}
\vspace{-0.2in}
\end{figure}

\begin{figure}[!t]
\centering
\includegraphics[width=0.7\textwidth]{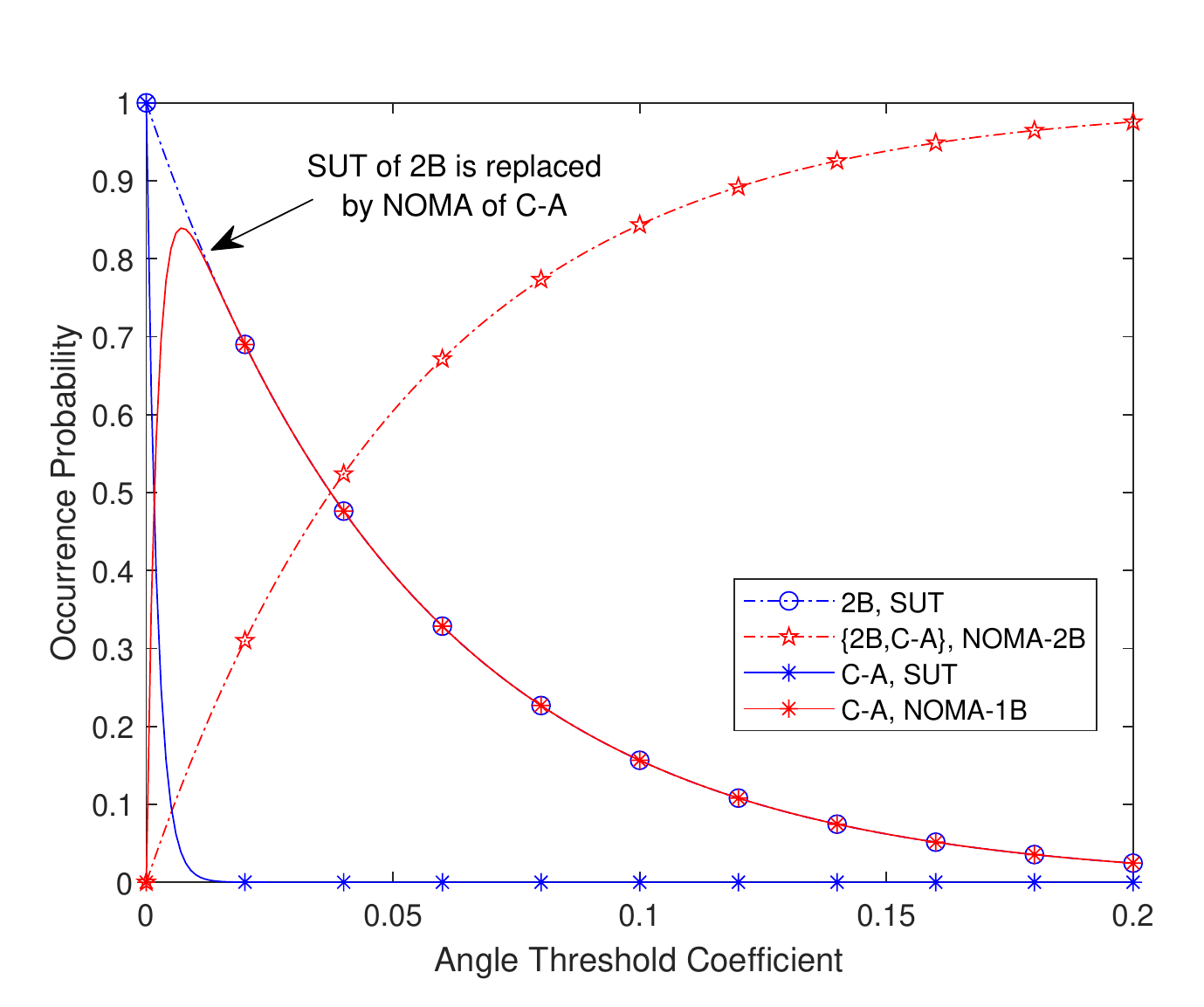}
\vspace{0.0in}
\caption{Occurrence probabilities for two-bit NOMA (2B) and combined NOMA with angle-only one-bit feedback with $d_\mathsf{max} \,{=}\, 27\,\text{m}$ and $\Delta \,{=}\, 10^\circ$.}
\label{fig:occurrence_dmax27_beam10}
\vspace{-0.2in}
\end{figure}

In Fig.~\ref{fig:geometry_bw15_dmax45}\subref{fig:geometry_cnoma_bw15_dmax45}, we include the combined NOMA schemes while finding the optimal strategy. We observe that some $(d_\mathsf{max},\Delta)$ pairs which are previously associated with either two-bit NOMA or angle-only (distance-only) one-bit NOMA in Fig.~\ref{fig:geometry_bw15_dmax45}\subref{fig:geometry_noma_bw15_dmax45} are now assigned to combined NOMA with angle-only (distance-only) one-bit feedback. To gain more insight into how this new assignment improves the performance, we present the hybrid sum-rate results for $(d_\mathsf{max},\Delta) \,{=}\, (27\,\text{m},10^\circ)$ in Fig.~\ref{fig:rate_dmax27_beam10}, for which the optimal strategy is one-bit NOMA and combined NOMA (both with angle-only feedback) in Fig.~\ref{fig:geometry_bw15_dmax45}\subref{fig:geometry_noma_bw15_dmax45} and Fig.~\ref{fig:geometry_bw15_dmax45}\subref{fig:geometry_cnoma_bw15_dmax45}, respectively. We observe in Fig.~\ref{fig:rate_dmax27_beam10} that the combined NOMA with angle-only feedback outperforms to other NOMA strategies, with a significant gap between either one-bit or two-bit NOMA. We also depict representative occurrence probabilities against varying angle threshold in Fig.~\ref{fig:occurrence_dmax27_beam10} for this particular setting (considering two-bit NOMA and combined NOMA with angle-only feedback). We observe how the occurrence probability for SUT associated with two-bit NOMA is replaced by that for angle-only one-bit NOMA of combined NOMA strategy. By this shift from---simply---SUT to NOMA, the hybrid sum rates associated with combined NOMA are significantly better than that of two-bit NOMA, as illustrated in Fig.~\ref{fig:rate_dmax27_beam10}.        

\section{Conclusion}\label{sec:conclusion}
We propose a practical NOMA transmission strategy for mmWave communications, which makes use of unique angular propagation characteristics of mmWave links. In particular, we propose a low-resolution feedback mechanism for NOMA, where the distance and angle information of UEs are represented by one bit each. The transmitter then splits UEs into groups of strong and weak channels based on this two-bit information, and sets up UE pairs for NOMA transmission. Whenever a UE pair is not possible through two-bit feedback, the overall transmission mechanism resorts either to a hybrid NOMA scheme using one-bit information individually, as well, or all the spectral resources are allocated to a single UE. The numerical results verify the superiority of the proposed NOMA strategy over the existing solutions.    

\appendices

\section{Proof of Theorem~\ref{theorem:cdf_ecg}} \label{appendix:cdf_ecg}

We start the derivation by assuming two-bit NOMA for which the CDF of the effective channel gain of the strong UE is given as follows  
\begin{align}
    \mathsf{F}_\mathsf{S}^{\mathsf{2B}}(x) 
    &= \mathsf{Pr} \left\lbrace  
    |\textbf{h}_{k}^{\rm H}\textbf{b}|^2 \leq x \mid k \in \mathcal{S}_\mathsf{S}^\mathsf{2B} 
    \right\rbrace , \label{eq:app:cdf_twobit_strong_1} \\
    &= \frac{\mathsf{Pr} \left\lbrace  
    |\textbf{h}_{k}^{\rm H}\textbf{b}|^2 \leq x , | \overline{\theta} \,{-}\, \theta_k| \leq \theta_\mathsf{th} , d_k \leq d_\mathsf{th} 
    \right\rbrace}
    {\mathsf{Pr} \left\lbrace  
    | \overline{\theta} \,{-}\, \theta_k| \leq \theta_\mathsf{th} , d_k \leq d_\mathsf{th} 
    \right\rbrace}. \label{eq:app:cdf_twobit_strong_2} 
\end{align}
Recall that the angle follows uniform distribution with the probability distribution function (PDF) $f_\theta(x) \,{=}\, 1/\Delta$ for $|\overline{\theta} \,{-}\, x| \,{\leq}\, \Delta/2$, and $0$ otherwise. In addition, the distance has the PDF given by $f_d(x) \,{=}\, 2x/(d_\mathsf{max}^2 \,{-}\, d_\mathsf{min}^2)$ for $d_\mathsf{min} \,{\leq}\, x \,{\leq}\, d_\mathsf{max}$, and $0$ otherwise, which is statistically independent of the angle \cite{Yapici2019AngFee}. Using the definition in \eqref{eq:effective_channel_gain}, the CDF in \eqref{eq:app:cdf_twobit_strong_2} yields
\begin{align}
    \mathsf{F}_\mathsf{S}^{\,\mathsf{2B}}(x) 
    &= \frac{\Delta \left( d_\mathsf{max}^2 - d_\mathsf{min}^2 \right) }
    {2\theta_\mathsf{th} \left( d_\mathsf{th}^2 - d_\mathsf{min}^2 \right)} 
    \int\limits_{d_\mathsf{min}}^{d_\mathsf{th}} \int\limits_{\overline{\theta} \,{-}\, \theta_\mathsf{th}}^{\overline{\theta} \,{+}\, \theta_\mathsf{th}} \mathsf{Pr} \left\lbrace  
   |\alpha_{k}|^2 \leq \frac{\mathsf{PL}(r)}{\mathsf{F}_M(\overline{\theta},\theta)} x \right\rbrace
   f_d(r) f_\theta(\theta) \dd \theta \dd r .
   \label{eq:app:cdf_twobit_strong_3} 
\end{align}
Since the complex channel gain representing the impact of small-scale fading follows Gaussian distribution, its power is exponential with the CDF of $\mathsf{F}_\alpha(x) \,{=}\, 1 \,{-}\, \exp \left\lbrace {-}\frac{x}{\sigma^2}\right\rbrace$ for $x \,{\geq}\, 0$, and $0$ otherwise. Expressing the probability in \eqref{eq:app:cdf_twobit_strong_3} using the function $\mathsf{F}_\alpha$, and employing the explicit expressions of PDF's $f_\theta$ and $f_d$ in the integration, we readily obtain \eqref{eq:cdf_ecg} after simple algebraic manipulations, and by using the definitions $\xi_\mathsf{S}^\mathsf{2B}$, $\mathcal{R}_d^\mathsf{S,2B}$, and $\mathcal{R}_\theta^\mathsf{S,2B}$ of Theorem~\ref{theorem:cdf_ecg}. Note that \eqref{eq:app:cdf_twobit_strong_3} is a pretty general description valid for any path-loss pattern $\mathsf{PL}(r)$, and is very complicated in terms of obtaining a closed-form expression due to the complexity of $\mathsf{F}_M(\overline{\theta},\theta)$. \hfill\IEEEQEDhere

Similarly, the CDF of the effective channel gain for the weak UE is  
\begin{align}
    \mathsf{F}_\mathsf{W}^{\,\mathsf{2B}}(x) 
    &= \mathsf{Pr} \left\lbrace  
    |\textbf{h}_{k}^{\rm H}\textbf{b}|^2 \leq x \mid k \in \mathcal{S}_\mathsf{W}^\mathsf{2B} 
    \right\rbrace , \label{eq:app:cdf_twobit_weak_1} \\
    &= \frac{\mathsf{Pr} \left\lbrace  
    |\textbf{h}_{k}^{\rm H}\textbf{b}|^2 \leq x , | \overline{\theta} \,{-}\, \theta_k| > \theta_\mathsf{th} , d_k > d_\mathsf{th} 
    \right\rbrace}
    {\mathsf{Pr} \left\lbrace  
    | \overline{\theta} \,{-}\, \theta_k| > \theta_\mathsf{th} , d_k > d_\mathsf{th} 
    \right\rbrace}, \label{eq:app:cdf_twobit_weak_2} 
\end{align}
which can be manipulated similar to \eqref{eq:app:cdf_twobit_strong_3} as follows
\begin{align}
    \mathsf{F}_\mathsf{W}^{\,\mathsf{2B}}(x) 
    &= \frac{\Delta \left( d_\mathsf{max}^2 - d_\mathsf{min}^2 \right) }
    {\left(\Delta - 2\theta_\mathsf{th} \right) \left( d_\mathsf{max}^2 - d_\mathsf{th}^2 \right)} 
    \Bigg [
    \int\limits_{d_\mathsf{th}}^{d_\mathsf{max}} 
    \int\limits_{\overline{\theta} \,{-}\, \frac{\Delta}{2}}^{\overline{\theta} \,{-}\, \theta_\mathsf{th}} \left( 1 - \exp \left\lbrace {-}\frac{\mathsf{PL}(r)}{\mathsf{F}_M(\overline{\theta},\theta )}\frac{x}{\sigma^2} \right\rbrace \right)
   f_d(r) f_\theta(\theta) \dd \theta \dd r \nonumber \\
   & \hspace{1in} + \int\limits_{d_\mathsf{th}}^{d_\mathsf{max}} 
    \int\limits_{\overline{\theta} \,{+}\, \theta_\mathsf{th}}^{\overline{\theta} \,{+}\, \frac{\Delta}{2}} \left( 1 - \exp \left\lbrace {-}\frac{\mathsf{PL}(r)}{\mathsf{F}_M(\overline{\theta},\theta )}\frac{x}{\sigma^2} \right\rbrace \right)
   f_d(r) f_\theta(\theta) \dd \theta \dd r \Bigg].
   \label{eq:app:cdf_twobit_weak_3} 
\end{align}
As before, after simple algebra and by the definitions of $\xi_\mathsf{W}^\mathsf{2B}$, $\mathcal{R}_d^\mathsf{W,2B}$, and $\mathcal{R}_\theta^\mathsf{W,2B}$ in Theorem~\ref{theorem:cdf_ecg}, we can represent \eqref{eq:app:cdf_twobit_weak_3} by \eqref{eq:cdf_ecg}. \hfill\IEEEQEDhere

Following a similar strategy, the CDF of the effective channel gain for the one-bit NOMA with the angle feedback can be obtained for the strong UE as follows 
\begin{align}
    \mathsf{F}_\mathsf{S}^{\mathsf{A}}(x) 
    &= \mathsf{Pr} \left\lbrace  
    |\textbf{h}_{k}^{\rm H}\textbf{b}|^2 \leq x \mid k \in \mathcal{S}_\mathsf{S}^\mathsf{A} 
    \right\rbrace = \frac{\mathsf{Pr} \left\lbrace  
    |\textbf{h}_{k}^{\rm H}\textbf{b}|^2 \leq x , | \overline{\theta} \,{-}\, \theta_k| \leq \theta_\mathsf{th}
    \right\rbrace}
    {\mathsf{Pr} \left\lbrace  
    | \overline{\theta} \,{-}\, \theta_k| \leq \theta_\mathsf{th}     \right\rbrace}, \label{eq:app:cdf_onebit_angle_1} \\
    &= \frac{\Delta}{2\theta_\mathsf{th}}
    \int\limits_{d_\mathsf{min}}^{d_\mathsf{max}} \int\limits_{\overline{\theta} \,{-}\, \theta_\mathsf{th}}^{\overline{\theta} \,{+}\, \theta_\mathsf{th}} \left( 1 - \exp \left\lbrace {-}\frac{\mathsf{PL}(r)}{\mathsf{F}_M(\overline{\theta},\theta )}\frac{x}{\sigma^2} \right\rbrace \right)
   f_d(r) f_\theta(\theta) \dd \theta \dd r , \label{eq:app:cdf_onebit_angle_2}
\end{align}
and that for the weak UE is given as
\begin{align}
    \mathsf{F}_\mathsf{W}^{\mathsf{A}}(x) 
    &= \mathsf{Pr} \left\lbrace  
    |\textbf{h}_{k}^{\rm H}\textbf{b}|^2 \leq x \mid k \in \mathcal{S}_\mathsf{W}^\mathsf{2B} 
    \right\rbrace = \frac{\mathsf{Pr} \left\lbrace  
    |\textbf{h}_{k}^{\rm H}\textbf{b}|^2 \leq x , | \overline{\theta} \,{-}\, \theta_k| > \theta_\mathsf{th}
    \right\rbrace}
    {\mathsf{Pr} \left\lbrace  
    | \overline{\theta} \,{-}\, \theta_k| > \theta_\mathsf{th}
    \right\rbrace}, \label{eq:app:cdf_onebit_angle_3} \\
    &= \frac{\Delta}{\Delta - 2\theta_\mathsf{th}} 
    \Bigg [
    \int\limits_{d_\mathsf{min}}^{d_\mathsf{max}} 
    \int\limits_{\overline{\theta} \,{-}\, \frac{\Delta}{2}}^{\overline{\theta} \,{-}\, \theta_\mathsf{th}} \left( 1 - \exp \left\lbrace {-}\frac{\mathsf{PL}(r)}{\mathsf{F}_M(\overline{\theta},\theta )}\frac{x}{\sigma^2} \right\rbrace \right)
   f_d(r) f_\theta(\theta) \dd \theta \dd r \nonumber \\
   & \hspace{1in} + \int\limits_{d_\mathsf{min}}^{d_\mathsf{max}} 
    \int\limits_{\overline{\theta} \,{+}\, \theta_\mathsf{th}}^{\overline{\theta} \,{+}\, \frac{\Delta}{2}} \left( 1 - \exp \left\lbrace {-}\frac{\mathsf{PL}(r)}{\mathsf{F}_M(\overline{\theta},\theta )}\frac{x}{\sigma^2} \right\rbrace \right)
   f_d(r) f_\theta(\theta) \dd \theta \dd r \Bigg],
   \label{eq:app:cdf_onebit_angle_4} 
\end{align}
where both \eqref{eq:app:cdf_onebit_angle_2} and \eqref{eq:app:cdf_onebit_angle_4} is equivalent to \eqref{eq:cdf_ecg} for $\mathsf{s} \,{\in}\, \{\mathsf{S,W}\}$ with the definitions $\xi_\mathsf{s}^\mathsf{A}$, $\mathcal{R}_d^\mathsf{s,A}$, and $\mathcal{R}_\theta^\mathsf{s,A}$. We omit the derivation of the desired CDF's for one-bit NOMA with distance feedback since it is very similar to angle-only feedback as described by \eqref{eq:app:cdf_onebit_angle_1}-\eqref{eq:app:cdf_onebit_angle_4}. \hfill\IEEEQEDhere   

\section{Proof of Theorem~\ref{theorem:event_prob_sut}} \label{appendix:event_prob_sut}

Assuming two-bit NOMA, the overall transmission mechanism switches to the SUT scheme whenever NOMA is not feasible (i.e., at least one of $\mathcal{S}_\mathsf{S}^\mathsf{2B}$ and $\mathcal{S}_\mathsf{W}^\mathsf{2B}$ is empty). When the scheduled UE is picked up from $\mathcal{S}_\mathsf{S}^\mathsf{2B}$, the occurrence probability of SUT for a given $K$ (the total number of UEs) is given as
\begin{align}
    \mathsf{P}_{\mathsf{SUT}|K}^\mathsf{S,2B} &= \mathsf{Pr} \left\lbrace  \left| \mathcal{S}_\mathsf{S}^\mathsf{2B} \right| {\geq} \, 1, \left| \mathcal{S}_\mathsf{W}^\mathsf{2B} \right| {=} 0 , | 
    \overline{\mathcal{S}}^\mathsf{2B} | \,{=}\, K \,{-}\,  \left| \mathcal{S}_\mathsf{S}^\mathsf{2B} \right|  \right\rbrace , \label{eq:app:sut_two_1} \\
    &= \sum_{n=1}^{K} \mathsf{Pr} \left\lbrace  \left| \mathcal{S}_\mathsf{S}^\mathsf{2B} \right| {=} \, n, \left| \mathcal{S}_\mathsf{W}^\mathsf{2B} \right| {=} 0 , | 
    \overline{\mathcal{S}}^\mathsf{2B} | \,{=}\, K \,{-}\,  n  \right\rbrace , \label{eq:app:sut_two_2}
\end{align}
where $\overline{\mathcal{S}}^\mathsf{2B}$ stands for the set of UEs present in neither $\mathcal{S}_\mathsf{S}^\mathsf{2B}$ nor $\mathcal{S}_\mathsf{W}^\mathsf{2B}$. 

Note that any UE belongs to the groups $\mathcal{S}_\mathsf{S}^\mathsf{2B}$, $\mathcal{S}_\mathsf{W}^\mathsf{2B}$, and $\overline{\mathcal{S}}^\mathsf{2B}$ with the probabilities $p_\mathsf{S,2B}$, $p_\mathsf{W,2B}$, and $\overline{p}_\mathsf{2B} \,{=}\, 1\,{-}\,p_\mathsf{S,2B}\,{-}\,p_\mathsf{W,2B}$, respectively. The presence of any UE in $\mathcal{S}_\mathsf{S}^\mathsf{2B}$, $\mathcal{S}_\mathsf{W}^\mathsf{2B}$, or $\overline{\mathcal{S}}^\mathsf{2B}$ can therefore be represented by Bernoulli distribution with the probabilities $p_\mathsf{S,2B}$, $p_\mathsf{W,2B}$, and $\overline{p}_\mathsf{2B}$, respectively. As a result, the number of UEs present in any of these three regions can be approximated by multinomial distribution through the sum of Bernoulli random variables \cite{Yapici2019NonVlc}, and \eqref{eq:app:sut_two_2} accordingly becomes 
\begin{align}
    \mathsf{P}_{\mathsf{SUT}|K}^\mathsf{S,2B}     &= \sum_{n=1}^{K} \binom{K}{n} p_\mathsf{S,2B}^n \left(1\,{-}\,p_\mathsf{S,2B}\,{-}\,p_\mathsf{W,2B}\right)^{K{-}n}, \label{eq:app:sut_two_3}
\end{align}
where $\binom{K}{n} \,{=}\, K!{/}n!(K{-}n)!$ is the binomial coefficient. Note that although the probability in \eqref{eq:app:sut_two_2} is computed using multinomial distribution, it ends up with binomial coefficient followed by two probability terms as in \eqref{eq:app:sut_two_3}, since we are looking for cases where $\mathcal{S}_\mathsf{W}^\mathsf{2B}$ is an empty set. 

By Binomial theorem, \eqref{eq:app:sut_two_3} can be expressed as 
\begin{align}
    \mathsf{P}_{\mathsf{SUT}|K}^\mathsf{S,2B} &= \left(1\,{-}\,p_\mathsf{W,2B}\right)^{K}- p_\mathsf{\overline{2B}}^K, \label{eq:app:sut_two_4}
\end{align}
and averaging \eqref{eq:app:sut_two_4} over the total number of UEs being Poisson produces
\begin{align}
    \mathsf{P}_{\mathsf{SUT}}^\mathsf{S,2B} &= \sum\limits_{K=1}^\infty \left[\left(1\,{-}\,p_\mathsf{W,2B}\right)^{K}- p_\mathsf{\overline{2B}}^K \right] {\rm e}^{{-}\mu} \frac{\mu^K}{K!}, \label{eq:app:sut_two_5} \\
    &= {\rm e}^{{-}\mu}  \left[ \sum\limits_{K=1}^\infty \frac{\left( \left[1\,{-}\,p_\mathsf{W,2B}\right]\mu\right)^{K} }{K!} -  \sum\limits_{K=1}^\infty \frac{\left(\mu p_\mathsf{\overline{2B}}\right)^K}{K!} \right], \label{eq:app:sut_two_6}
\end{align}
which yields \eqref{eq:prob_sut_twobit} after employing Taylor series expansion ${\rm e}^x \,{=}\, \sum_{n=0}^\infty x^n/n!$ and simple algebra. Note that SUT is possible for any $K \,{\geq}\, 1$, and hence the summation in \eqref{eq:app:sut_two_5} considers any applicable $K$ value. Similarly, when a UE from  $\mathcal{S}_\mathsf{W}^\mathsf{2B}$ is scheduled, we have
\begin{align}
    \mathsf{P}_{\mathsf{SUT}|K}^\mathsf{W,2B} &= \mathsf{Pr} \left\lbrace  \left| \mathcal{S}_\mathsf{S}^\mathsf{2B} \right| {=} \, 0, \left| \mathcal{S}_\mathsf{W}^\mathsf{2B} \right| {\geq} 1 , | 
    \overline{\mathcal{S}}^\mathsf{2B} | \,{=}\, K \,{-}\,  \left| \mathcal{S}_\mathsf{W}^\mathsf{2B} \right|  \right\rbrace , \label{eq:app:sut_two_7} \\
    &= \sum_{n=1}^{K} \binom{K}{n} p_\mathsf{W,2B}^n \left(1\,{-}\,p_\mathsf{S,2B}\,{-}\,p_\mathsf{W,2B}\right)^{K{-}n},  \label{eq:app:sut_two_8}  \\
    &= \left(1\,{-}\,p_\mathsf{S,2B}\right)^{K}- p_\mathsf{\overline{2B}}, \label{eq:app:sut_two_9}
\end{align}
which also yields \eqref{eq:prob_sut_twobit} following a similar strategy. \hfill\IEEEQEDhere

For one-bit NOMA with either the angle- or distance-only feedback, the complete set of all the UEs is comprised of set of strong and weak UEs, and hence $\overline{\mathcal{S}}^\mathsf{t} \,{=}\, \varnothing$ for $\mathsf{t} \,{\in}\, \{\mathsf{A,D}\}$. In other words, $p_\mathsf{S,t}\,{+}\,p_\mathsf{W,t} \,{=}\, 1$, and hence $\overline{p}_\mathsf{2B} \,{=}\, 0$. As a result, the number of UEs in either $\mathcal{S}_\mathsf{S}^\mathsf{t}$ or $\mathcal{S}_\mathsf{S}^\mathsf{t}$ follows binomial distribution, and the respective SUT occurrence probabilities are given as
\begin{align}
    \mathsf{P}_{\mathsf{SUT}|K}^\mathsf{S,t} &= \mathsf{Pr} \left\lbrace  \left| \mathcal{S}_\mathsf{S}^\mathsf{t} \right| {=} \, K, \left| \mathcal{S}_\mathsf{W}^\mathsf{t} \right| {=} \, 0 \right\rbrace = p_\mathsf{S,t}^K , \label{eq:app:sut_one_1} \\
    \mathsf{P}_{\mathsf{SUT}|K}^\mathsf{W,t} &= \mathsf{Pr} \left\lbrace  \left| \mathcal{S}_\mathsf{S}^\mathsf{t} \right| {=} \, 0, \left| \mathcal{S}_\mathsf{W}^\mathsf{t} \right| {=} \, K \right\rbrace = p_\mathsf{W,t}^K , \label{eq:app:sut_one_2}
\end{align}
which yield \eqref{eq:prob_sut_one} after following the same strategy applied to \eqref{eq:app:sut_two_4} while obtaining \eqref{eq:prob_sut_twobit}. Note also the similarity of \eqref{eq:prob_sut_twobit} and \eqref{eq:prob_sut_one} for $\overline{p}_\mathsf{2B} \,{=}\, 0$, which holds for one-bit feedback.    \hfill\IEEEQEDhere

\section{Proof of Theorem~\ref{theorem:event_prob_noma}} \label{appendix:event_prob_noma}

The NOMA transmission of any type occurs whenever there is at least one UE available in both strong and weak UE groups. The respective occurrence probability for a given $K$ (the total number of UEs) can be therefore given for any feedback type as follows  
\begin{align}
    \mathsf{P}_{\mathsf{NOMA}|K}^\mathsf{t} \,{=}\, \mathsf{Pr} \left\lbrace  \left| \mathcal{S}_\mathsf{S}^\mathsf{t} \right| {\geq} \, 1, \left| \mathcal{S}_\mathsf{W}^\mathsf{t} \right| {\geq} \, 1, | 
    \overline{\mathcal{S}}^\mathsf{t} | \,{=}\, K \,{-}\,  \left| \mathcal{S}_\mathsf{S}^\mathsf{t} \right|  \,{-}\,  \left| \mathcal{S}_\mathsf{W}^\mathsf{t} \right|   \right\rbrace , \label{eq:app:noma_prob_1}
\end{align}
with $\mathsf{t} \,{\in}\, \{\mathsf{2B,A,D}\}$. For two-bit NOMA, \eqref{eq:app:noma_prob_1} becomes
\begin{align}
    \mathsf{P}_{\mathsf{NOMA}|K}^\mathsf{2B} &= \sum_{n=1}^{K-1} \mathsf{Pr} \left\lbrace  \left| \mathcal{S}_\mathsf{S}^\mathsf{2B} \right| {=} \, n, \left| \mathcal{S}_\mathsf{W}^\mathsf{2B} \right| {\geq} \, 1, | 
    \overline{\mathcal{S}}^\mathsf{2B} | \,{=}\, K \,{-}\,  n  \,{-}\,  \left| \mathcal{S}_\mathsf{W}^\mathsf{2B} \right|   \right\rbrace , \label{eq:app:noma_prob_2} \\
    &= \sum_{n=1}^{K-1} \sum_{m=1}^{K-1-n} \mathsf{Pr} \left\lbrace  \left| \mathcal{S}_\mathsf{S}^\mathsf{2B} \right| {=} \, n, \left| \mathcal{S}_\mathsf{W}^\mathsf{2B} \right| {=} \, m, | 
    \overline{\mathcal{S}}^\mathsf{2B} | \,{=}\, K \,{-}\,  n  \,{-}\, m \right\rbrace , \label{eq:app:noma_prob_3}
\end{align}
where the probability in \eqref{eq:app:noma_prob_3} can be represented using the multinomial distribution as follows
\begin{align}
    \mathsf{P}_{\mathsf{NOMA}|K}^\mathsf{2B} &= \sum_{n=1}^{K-1} \sum_{m=1}^{K-1-n} \frac{(K{-}1)!}{n!m!(K{-}1{-}n{-}m)!} p_\mathsf{S,2B}^n \,  p_\mathsf{W,2B}^m \,  \left(1\,{-}\,p_\mathsf{S,2B}\,{-}\,p_\mathsf{W,2B}\right)^{K{-}n{-}m} . \label{eq:app:noma_prob_4}
\end{align}
Although the desired occurrence probability can be obtained by averaging \eqref{eq:app:noma_prob_4} over all possible $K$ values, which follow Poisson distribution, the result would be computationally inefficient due to the multiple summation operators.

Note that two-bit feedback might end up with a no-transmission situation as well as the respective NOMA and SUT schemes. An alternative way to find $\mathsf{P}_\mathsf{NOMA}^\mathsf{2B}$ is therefore consider all possible scenarios associated with two-bit feedback as follows
\begin{align}\label{eq:app:all}
    \mathsf{P}_\mathsf{NOMA}^\mathsf{2B} = 1 - \mathsf{P}_\mathsf{SUT}^\mathsf{S,2B} - \mathsf{P}_\mathsf{SUT}^\mathsf{W,2B} - \mathsf{P}_\mathsf{NO}^\mathsf{2B}, 
\end{align}
where $\mathsf{P}_\mathsf{NO}^\mathsf{2B}$ is the respective occurrence probability of no-transmission cases, and is given as
\begin{align}
    \mathsf{P}_\mathsf{NO}^\mathsf{2B} &= \sum_{K=0}^\infty \mathsf{Pr} \left\lbrace  \left| \mathcal{S}_\mathsf{S}^\mathsf{2B} \right| {=} \, 0, \left| \mathcal{S}_\mathsf{W}^\mathsf{2B} \right| {=} 0 , | 
    \overline{\mathcal{S}}^\mathsf{2B} | \,{=}\, K   \right\rbrace 
    {\rm e}^{{-}\mu} \frac{\mu^K}{K!} =  {\rm e}^{{-}\mu\left(1-p_\mathsf{\overline{2B}}\right)}. \label{eq:app:no}
\end{align}
Finally employing \eqref{eq:prob_sut_twobit} and \eqref{eq:app:no} in \eqref{eq:app:all} yields \eqref{eq:prob_noma_twobit}. \hfill\IEEEQEDhere

A similar approach for one-bit NOMA produces the desired occurrence probability as follows 
\begin{align}
    \mathsf{P}_{\mathsf{NOMA}|K}^\mathsf{t} &= \sum_{n=1}^{K-1} \mathsf{Pr} \left\lbrace  \left| \mathcal{S}_\mathsf{S}^\mathsf{t} \right| {=} \, n, \left| \mathcal{S}_\mathsf{W}^\mathsf{t} \right| {=} \, K \,{-}\,  n   \right\rbrace = \sum_{n=1}^{K-1} \binom{K}{n} p_\mathsf{S,t}^n \,  p_\mathsf{W,t}^{K{-}n} , \label{eq:app:noma_prob_5}
\end{align}
with $\mathsf{t} \,{\in}\, \{\mathsf{A,D}\}$, which can be modified using Binomial theorem as follows
\begin{align}
    \mathsf{P}_{\mathsf{NOMA}|K}^\mathsf{t} &= 1 - p_\mathsf{S,t}^K - p_\mathsf{W,t}^K .  \label{eq:app:noma_prob_6}
\end{align}
The desired probability in \eqref{eq:prob_noma_onebit} is obtained readily following the same strategy applied to \eqref{eq:app:sut_two_4} while obtaining \eqref{eq:prob_sut_twobit}. \hfill\IEEEQEDhere

\section{Proof of Theorem~\ref{theorem:prob_sut_comb}} \label{appendix:prob_sut_comb}

For the case of combined NOMA transmission, the SUT scheme considers the order $\mathcal{S}_\mathsf{S}^\mathsf{2B}\rightarrow \mathcal{S}_\mathsf{S}^\mathsf{t} \rightarrow \mathcal{S}_\mathsf{W}^\mathsf{t}$ to pick up a UE to schedule if none of one-bit and two-bit NOMA is possible, with $\mathsf{t} \,{\in}\, \{\mathsf{A,D}\}$. We first consider the angle-based one-bit feedback to derive the respective SUT probability where the NOMA part employs two-bit and angle-based one-bit feedback schemes while the SUT part processes the order $\mathcal{S}_\mathsf{S}^\mathsf{2B}\rightarrow \mathcal{S}_\mathsf{S}^\mathsf{A} \rightarrow \mathcal{S}_\mathsf{W}^\mathsf{A}$. Under this condition, the SUT scheme finds at least one UE in $\mathcal{S}_\mathsf{S}^\mathsf{2B}$ when the scenarios 9 and 10 of Table~\ref{table:user_deployment} occurs, and the respective probability is given as
\begin{align}
    \mathsf{P}_{\mathsf{C-SUT}|K}^\mathsf{S,2B,A} &= \mathsf{Pr} \left\lbrace  \left| \mathcal{S}_\mathsf{S}^\mathsf{2B} \right| {\geq} \, 1, \left|    \mathcal{S}_{\overline{W}}^\mathsf{2B} \right| \,{=}\, K \,{-}\,  \left| \mathcal{S}_\mathsf{S}^\mathsf{2B} \right|, \left| \mathcal{S}_\mathsf{W}^\mathsf{2B} \right| = \left| \mathcal{S}_\mathsf{\overline{S}}^\mathsf{2B} \right| = 0   \right\rbrace , \label{eq:app:sut_comb_1} \\
    &= \sum_{n=1}^{K} \left\lbrace  \left| \mathcal{S}_\mathsf{S}^\mathsf{2B} \right| {=} \, n, \left|    \mathcal{S}_{\overline{W}}^\mathsf{2B} \right| \,{=}\, K \,{-}\,  n, \left| \mathcal{S}_\mathsf{W}^\mathsf{2B} \right| = \left| \mathcal{S}_\mathsf{\overline{S}}^\mathsf{2B} \right| = 0   \right\rbrace , \label{eq:app:sut_comb_2} \\
    &= \sum_{n=1}^{K} \binom{K}{n} p_\mathsf{S,2B}^n \, p_\mathsf{\overline{w}}^{K{-}n} = \left( p_\mathsf{S,2B} + p_\mathsf{\overline{w}} \right)^K - p_\mathsf{\overline{w}}^K, \label{eq:app:sut_comb_3}
\end{align}
where $\mathcal{S}_{\overline{\mathsf{S}}}^\mathsf{2B}$ and $\mathcal{S}_{\overline{\mathsf{W}}}^\mathsf{2B}$ are the group of UEs having specific bounds for their distance and angle values as sketched in Fig.~\ref{fig:setting}, and $p_\mathsf{\overline{w}} \,{=}\, p_\theta(1{-}p_d)$. Similar to the previous derivations, \eqref{eq:prob_sut_comb_two} is obtained readily after averaging \eqref{eq:app:sut_comb_3} over $K$ and by some algebraic manipulations. \hfill\IEEEQEDhere 

Furthermore, the probability of finding SUT UE in $\mathcal{S}_\mathsf{S}^\mathsf{A}$, which corresponds to the scenario 2 of Table~\ref{table:user_deployment}, is given as
\begin{align}
    \mathsf{P}_{\mathsf{SUT}|K}^\mathsf{S,A} &= \mathsf{Pr} \left\lbrace  \left| \mathcal{S}_\mathsf{\overline{W}}^\mathsf{2B} \right| {=} \, K, \left| \mathcal{S}_\mathsf{S}^\mathsf{2B} \right| = \left| \mathcal{S}_\mathsf{W}^\mathsf{2B} \right| = \left| \mathcal{S}_\mathsf{\overline{S}}^\mathsf{2B} \right| = 0   \right\rbrace = p_\mathsf{\overline{w}}^{K}, \label{eq:app:sut_comb_4}
\end{align}
and the desired probability of scheduling a UE from $\mathcal{S}_\mathsf{W}^\mathsf{A}$ is
\begin{align}
    \mathsf{P}_{\mathsf{SUT}|K}^\mathsf{W,A} &= \mathsf{Pr} \left\lbrace  \left| \mathcal{S}_\mathsf{W}^\mathsf{2B} \right| {\geq} \, 0, \left|    \mathcal{S}_{\overline{S}}^\mathsf{2B} \right| \,{=}\, K \,{-}\,  \left| \mathcal{S}_\mathsf{W}^\mathsf{2B} \right|, \left| \mathcal{S}_\mathsf{S}^\mathsf{2B} \right| = \left| \mathcal{S}_\mathsf{\overline{W}}^\mathsf{2B} \right| = 0   \right\rbrace , \label{eq:app:sut_comb_5} \\
    &= \sum_{n=0}^{K} \binom{K}{n} p_\mathsf{W,2B}^n \, p_\mathsf{\overline{s}}^{K{-}n} = \left( p_\mathsf{W,2B} + p_\mathsf{\overline{s}} \right)^K, \label{eq:app:sut_comb_6}
\end{align}
which follows from the scenarios 3, 5, and 7 of Table~\ref{table:user_deployment}, with $p_\mathsf{\overline{s}} \,{=}\, p_d(1{-}p_\theta)$. As before, we readily obtain \eqref{eq:prob_sut_comb_strong} and \eqref{eq:prob_sut_comb_weak} after averaging \eqref{eq:app:sut_comb_4} and \eqref{eq:app:sut_comb_6}, respectively, over $K$, and carrying out some algebraic manipulations. \hfill\IEEEQEDhere

We now consider the distance-based one-bit feedback such that the NOMA part employs two-bit and distance-based one-bit feedback schemes, and the SUT part processes the order $\mathcal{S}_\mathsf{S}^\mathsf{2B}\rightarrow \mathcal{S}_\mathsf{S}^\mathsf{D} \rightarrow \mathcal{S}_\mathsf{W}^\mathsf{D}$. The probability for the SUT UE being in $\mathcal{S}_\mathsf{S}^\mathsf{2B}$, which corresponds to the scenarios 9 and 11 of Table~\ref{table:user_deployment}, is now given as
\begin{align}
    \mathsf{P}_{\mathsf{SUT}|K}^\mathsf{S,2B,D} &= \mathsf{Pr} \left\lbrace  \left| \mathcal{S}_\mathsf{S}^\mathsf{2B} \right| {\geq} \, 1, \left|    \mathcal{S}_{\overline{S}}^\mathsf{2B} \right| \,{=}\, K \,{-}\,  \left| \mathcal{S}_\mathsf{S}^\mathsf{2B} \right|, \left| \mathcal{S}_\mathsf{W}^\mathsf{2B} \right| = \left| \mathcal{S}_\mathsf{\overline{W}}^\mathsf{2B} \right| = 0   \right\rbrace , \label{eq:app:sut_comb_7} \\
    &= \sum_{n=1}^{K} \binom{K}{n} p_\mathsf{S,2B}^n \, p_\mathsf{\overline{s}}^{K{-}n} = \left( p_\mathsf{S,2B} + p_\mathsf{\overline{s}} \right)^K - p_\mathsf{\overline{s}}^K. \label{eq:app:sut_comb_8}
\end{align}

Similarly, the probability of finding SUT UE in $\mathcal{S}_\mathsf{S}^\mathsf{D}$ (i.e., scenario 3 of Table~\ref{table:user_deployment}) is
\begin{align}
    \mathsf{P}_{\mathsf{SUT}|K}^\mathsf{S,D} &= \mathsf{Pr} \left\lbrace  \left| \mathcal{S}_\mathsf{\overline{S}}^\mathsf{2B} \right| {=} \, K, \left| \mathcal{S}_\mathsf{S}^\mathsf{2B} \right| = \left| \mathcal{S}_\mathsf{W}^\mathsf{2B} \right| = \left| \mathcal{S}_\mathsf{\overline{W}}^\mathsf{2B} \right| = 0   \right\rbrace = p_\mathsf{\overline{s}}^{K}, \label{eq:app:sut_comb_9}
\end{align}
and in $\mathcal{S}_\mathsf{W}^\mathsf{D}$ (i.e., scenarios 2, 5, and 6 of Table~\ref{table:user_deployment}) is
\begin{align}
    \mathsf{P}_{\mathsf{SUT}|K}^\mathsf{W,D} &= \mathsf{Pr} \left\lbrace  \left| \mathcal{S}_\mathsf{W}^\mathsf{2B} \right| {\geq} \, 0, \left|    \mathcal{S}_{\overline{W}}^\mathsf{2B} \right| \,{=}\, K \,{-}\,  \left| \mathcal{S}_\mathsf{W}^\mathsf{2B} \right|, \left| \mathcal{S}_\mathsf{S}^\mathsf{2B} \right| = \left| \mathcal{S}_\mathsf{\overline{S}}^\mathsf{2B} \right| = 0   \right\rbrace , \label{eq:app:sut_comb_10} \\
    &= \sum_{n=0}^{K} \binom{K}{n} p_\mathsf{W,2B}^n \, p_\mathsf{\overline{w}}^{K{-}n} = \left( p_\mathsf{W,2B} + p_\mathsf{\overline{w}} \right)^K. \label{eq:app:sut_comb_11}
\end{align}
As before, summing up \eqref{eq:app:sut_comb_8}, \eqref{eq:app:sut_comb_9}, and \eqref{eq:app:sut_comb_11}, and averaging over the total number of UEs, we yield \eqref{eq:prob_sut_comb_two}, \eqref{eq:prob_sut_comb_strong}, and \eqref{eq:prob_sut_comb_weak}, respectively. \hfill\IEEEQEDhere

\bibliographystyle{IEEEtran}
\bibliography{IEEEabrv,references}

\end{document}